# Analytic Theory and cQED Implementation of a Two-Qubit Refrigerator: Sub-100 mK Cavity Cooling from a 4 K Bath

Daryoosh Vashaee[a,b*] and Jahanfar Abouie[c]
[a] Department of Materials Science & Engineering, North Carolina State University, Raleigh, NC 27606, USA
[b] Department of Electrical & Computer Engineering, North Carolina State University, Raleigh, NC 27606, USA
[c] Department of Physics, Institute for Advanced Studies in Basic Sciences, Zanjan 45137-66731, Iran

## Abstract

We develop an experimentally grounded theory for cooling a phonon-coupled microwave cavity far below its ambient bath using a resettable two-qubit reservoir. Motivated by cQED hardware that can operate at the 1-4 K stage, we model a high-Q 3D cavity repeatedly and weakly engaged by internally correlated two-level pairs that are re-prepared and actively reset between interactions. From a Lindblad master equation for a cavity subject to both a phonon bath and a Poisson train of finite-duration ancilla interactions, we obtain closed-form steady states for the cavity photon number and effective temperature in two geometries: (i) only one atom of each pair couples to the cavity, and (ii) both atoms couple collectively. The one-atom geometry cools the cavity below the phonon bath but not below the pair's own temperature. In contrast, when both atoms couple, intra-pair coherence renormalizes the upward/downward transition rates seen by the cavity, creating a quantum-enhanced refrigeration effect that can drive the cavity well below the reservoir temperature when phonon damping is weak. Our detuning-aware, time-window-aware collision model (including finite interaction time, arrival rate, and cavity damping) yields analytic lineshapes and identifies broad "cooling valleys" near resonance, together with the crossover between reservoir-dominated and phonon-dominated regimes. Guided by these results, we outline a concrete cQED implementation with two transmons in a 3D cavity using nanosecond flux tuning and fast dissipative or measurement-based resets. Under realistic parameters (MHz-rate cycles, small-angle exchanges, measured reset fidelities), the engineered reservoir reaches $\approx$ 50 mK even when the ambient bath is 4 K. Together, these results provide a compact, analyzable route to sub-100 mK microwave modes within a multi-kelvin cryostat, opening a realistic avenue toward scalable, higher-temperature quantum hardware.

**Keywords:** Quantum refrigeration, Circuit QED, Superconducting qubits, Collision models, Quantum thermodynamics

## Introduction

The growth of quantum information science continues to strain cryogenic infrastructure: cooling power falls by orders of magnitude between the 1-4 K stage and the millikelvin base of a dilution refrigerator, yet today's superconducting qubits, high-Q microwave resonators, and semiconductor spin qubits are typically concentrated at the coldest plate where thermal budgets and wiring heat loads severely limit scale.[1-9] A complementary strategy to "moving all hardware colder" is to engineer local quantum reservoirs that impose favorable detailed balance on selected degrees of freedom (specific cavity modes, qubits, or mesoscopic devices) so they equilibrate to effective temperatures far below that of their host stage. Collision models, micromaser physics, and quantum thermal machines show how repeated interactions with few-level ancillae can produce such nontrivial steady states.[10-18] In particular, the energetics-of-correlations ("photo-Carnot") model of ref [19] demonstrates, at the level of principle, that a stream of correlated two-level systems can shift a cavity's effective temperature via correlation-modified transition rates.

Recent experiments and modeling with optically polarized solid-state spins have sharpened the practical appeal of mode refrigeration. Using nitrogen-vacancy (NV) ensembles pumped at room temperature, theory predicts microwave mode occupations corresponding to ≈ 116 K (≈ 261 photons) under realistic maser-style parameters, together with room-temperature cQED signatures across weak-to-strong coupling.[20] Continuous, steady-state demonstrations now reach ≈ 150 K for a 2.87-GHz mode by coupling a driven NV ensemble to a dielectric resonator and reading out the associated noise reduction; the observed behavior is captured by a linearized Tavis–Cummings model[21] of a “spin refrigerator”.[22] These room-temperature works establish that engineered reservoirs can cool an electromagnetic mode substantially below ambient and that simple, analyzable open-system descriptions suffice to guide design.[22]

* dvashae@ncsu.edu

Our focus is a distinct, but experimentally natural, regime tailored to cQED hardware at the 1-4 K stage. Instead of a macroscopic spin ensemble driven by optical pumping at 300 K, we consider a minimal ancilla composed of a pair of superconducting qubits repeatedly prepared, coupled to a high-Q cavity, and actively reset. Two ingredients separate our treatment from both the photo-Carnot idealization and the NV-ensemble experiments. First, we include a realistic phonon (thermal) bath coupled to the cavity at temperature $\mathrm{T_{bath}} \approx 1 - 4\,\mathrm{K}$ and derive closed-form birth-death equations that capture the competition between the engineered ancilla reservoir and the ambient phonon tether. Second, we analyze two interaction geometries that are natural in cQED: (i) a one-subsystem configuration where only one member of each prepared pair couples to the cavity, and (ii) a two-subsystem configuration where both qubits couple during the collision window. In the latter case, intra-pair quantum correlations enter the effective upward and downward rates and can bias detailed balance so that the cavity’s steady state drops below the temperature associated with the ancilla populations, an effect that persists even when the cavity is phonon-tethered, provided the engineered exchange dominates.

The physical setup and the two interaction geometries considered in this work are illustrated schematically in Figure 1. It shows the phonon-coupled cavity refrigerator studied here and the two interaction geometries considered below: a one-atom coupling configuration, in which only one atom of each correlated pair interacts with the cavity, and a two-atom coupling configuration, in which both atoms couple to the cavity.

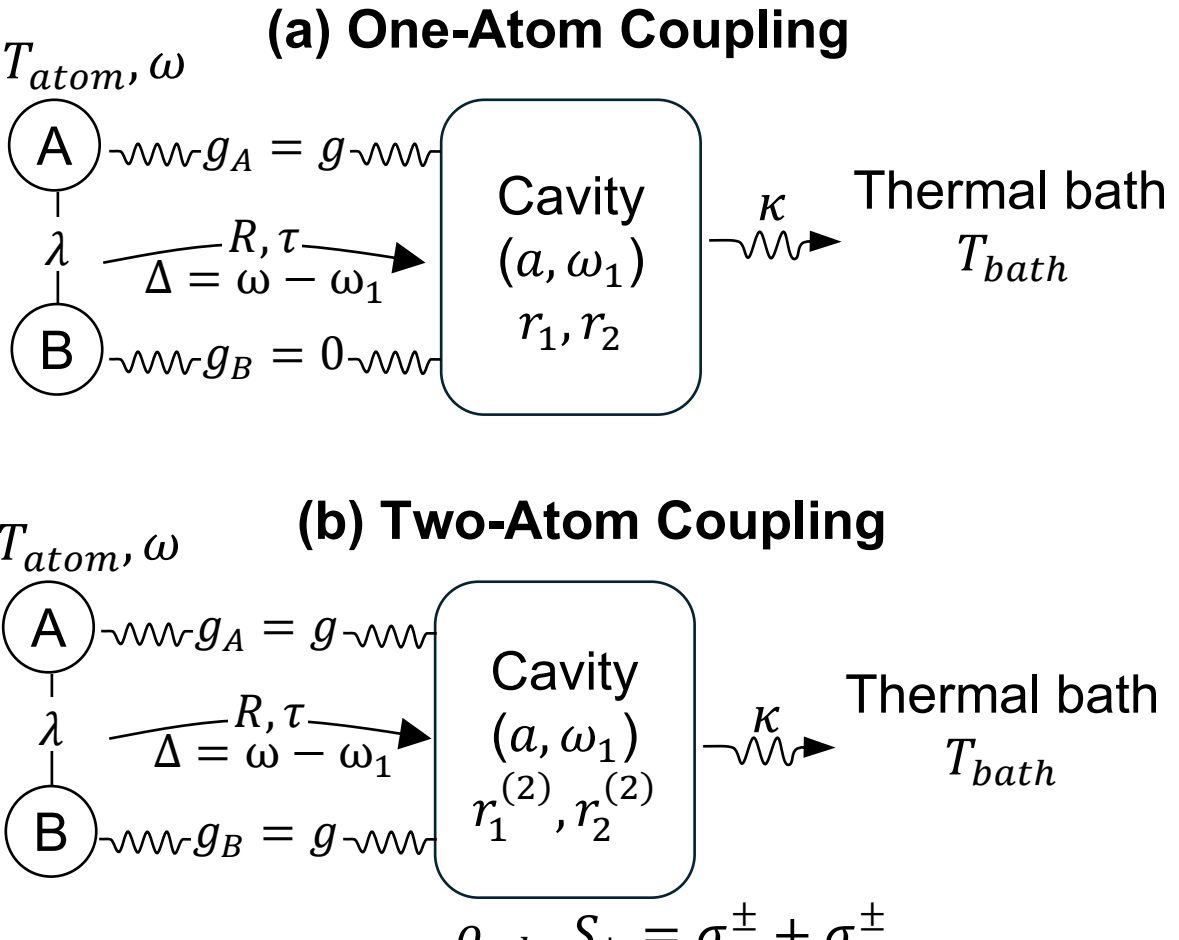

Figure 1: Schematic of the phonon-coupled cavity refrigerator. In both panels, a cavity mode of frequency $\omega_1$ is coupled to a thermal bath at temperature $T_{\mathrm{bath}}$ with damping rate $\kappa$, while correlated atom pairs characterized by $T_{\mathrm{atom}}$, $\omega$, $\lambda$, and arrival rate $R$ interact with the cavity for a duration $\tau$ at detuning $\Delta$. (a) One-atom coupling configuration: only atom $A$ couples to the cavity, with $g_A = g$ and $g_B = 0$. (b) Two-atom coupling configuration: both atoms couple to the cavity, $g_A = g_B = g$, allowing the pair coherence $\rho_{nd}$ to affect the effective transition rates.

Technically, we analyze short, weak, repeated interactions and employ a detuning-aware coarse graining that yields analytic steady-state photon numbers and effective temperatures as explicit functions of the atom–cavity coupling g, the interaction time $\tau$ (via $\phi = g\tau$), the arrival rate $\mathrm{R}$, the detuning $\Delta$ via a spectral-overlap filter, the cavity–bath damping $\kappa$ and an intra-pair exchange that controls both populations and coherence. The resulting closed-form birth-death equations identify reservoir-dominated versus phonon-dominated regimes, predict broad cooling valleys near resonance, and provide transparent conditions for achieving $T_{cav} < T_{bath}$ and, in the two-subsystem geometry, $T_{cav} < T_{atom}$. Conceptually, this furnishes a minimal, few-qubit, actively resettable reservoir native to cQED and amenable to MHz-rate prepare-interact-reset cycles, offering a complementary pathway to the ensemble-based, optically pumped mode refrigerators explored at room temperature.[20,22]

Finally, we map the theory to a concrete cQED implementation in which two transmons (or flux-type qubits) inside a 3D superconducting cavity are repeatedly prepared (including correlated mixed states), tuned into and out of resonance on nanosecond timescales, and autonomously or measurement-reset to maintain a low effective ancilla temperature. In this architecture the cavity is explicitly phonon-tethered to a 1-4 K environment, yet our analytic solutions show broad parameter regions where the engineered reservoir cools the cavity to the sub-100-mK range, well beyond what a passive 1-4 K stage would allow. Relative to the NV-ensemble demonstrations at room temperature,[20,22] our approach dispenses with optical pumping and large ensembles, leverages mature qubit reset and tunable coupling in cQED, and provides closed-form performance metrics tailored to phonon-coupled millikelvin targets embedded within higher-temperature stages. The qubits repeatedly interact with a cavity mode and are then actively reset using experimentally demonstrated protocols such as Purcell-enhanced dissipation, measurement-and-feedback reset, or autonomous reservoir engineering.[23-30] The cavity is additionally coupled to a phonon (or thermal) bath at temperature $T_{\text{bath}}$, which we take to lie around 1 K to reflect realistic high-power stages in modern cryogenic systems.[1-6] Within this framework, the qubit pair acts as a sacrificial low-entropy working medium, while the cavity is the object to be cooled; the resulting dynamics are naturally described by a collision model with tunable arrival rate, interaction time, detuning, and qubit–qubit correlations.[11]

## 1. Correlated atom-pair stream: only one atom coupled to a phonon-tethered cavity

### 1.1 System and parameters

We consider a single bosonic cavity mode $a$ of frequency $\omega_1$, described by the Hamiltonian

$$H_c = \hbar\omega_1 a^\dagger a$$

The cavity is coupled to an ambient phonon bath at temperature $T_{\text{bath}}$, which we model phenomenologically as a thermal Lindblad contact resonant at $\omega_1$ with overall damping rate $\kappa$. The corresponding Bose–Einstein occupation of the bath at the cavity frequency is

$$\bar{n}_1 = \frac{1}{\exp\left(\frac{\hbar\omega_1}{k_B T_{\text{bath}}}\right) - 1} \qquad (1)$$

In addition to its coupling to the thermal phonon environment, the cavity is repeatedly and weakly perturbed by a stream of correlated pairs of two-level atoms, which act as an engineered reservoir. Each incoming pair is prepared in the thermal state of an $XY$-type Heisenberg Hamiltonian with local level splitting $\hbar\omega$ and intra-pair exchange coupling $\hbar\lambda$. Writing $\beta \equiv 1/(k_B T_{\text{atom}})$, the pair density operator $\rho_{\text{pair}}$ is most conveniently expressed in the product basis $\{|ee\rangle, |eg\rangle, |ge\rangle, |gg\rangle\}$. In this basis, the relevant matrix elements are

$$\rho_e \equiv \langle ee \mid \rho_{\text{pair}} \mid ee \rangle, \qquad \rho_g \equiv \langle gg \mid \rho_{\text{pair}} \mid gg \rangle,$$

$$\rho_d \equiv \langle eg \mid \rho_{\text{pair}} \mid eg \rangle = \langle ge \mid \rho_{\text{pair}} \mid ge \rangle, \qquad \rho_{nd} \equiv \langle eg \mid \rho_{\text{pair}} \mid ge \rangle = \langle ge \mid \rho_{\text{pair}} \mid eg \rangle$$

For the thermal pair state considered here, these matrix elements take the form[19,17]

$$\rho_e = \frac{e^{-\beta\hbar\omega}}{Z}, \qquad \rho_g = \frac{e^{+\beta\hbar\omega}}{Z}, \qquad \rho_d = \frac{\cosh(\beta\hbar\lambda)}{Z}, \qquad \rho_{nd} = -\frac{\sinh(\beta\hbar\lambda)}{Z}, \qquad (2)$$

with partition function

$$Z = 2[\cosh(\beta\hbar\omega) + \cosh(\beta\hbar\lambda)] \qquad (3)$$

During a short interaction window of duration $\tau$, only one member of each pair, denoted atom $A$, is coupled to the cavity. The interaction is described by the Jaynes–Cummings Hamiltonian

$$H_{\text{int}} = \hbar g(a\,\sigma_A^+ + a^\dagger \sigma_A^-), \qquad (4)$$

where $\sigma_A^\pm$ are the ladder operators of atom $A$. We work throughout in the weak, short-collision limit

$$\phi \equiv g\tau \ll 1, \qquad (5)$$

so that each passage produces only a small perturbative "kick" to the cavity state. The incoming pairs are assumed to arrive independently as a Poisson process with average rate $R$, which ensures memoryless collisions on the coarse-grained timescale.

Because only atom $A$ couples to the field, the cavity dynamics depend only on the reduced state of that atom. It is therefore useful to define the effective one-atom stream coefficients directly as the local excited- and ground-state weights of atom $A$,

$$r_1 \equiv \text{Tr}\big(\sigma_A^+ \sigma_A^-\, \rho_{\text{pair}}\big) = \rho_e + \rho_d, \qquad r_2 \equiv \text{Tr}\big(\sigma_A^- \sigma_A^+\, \rho_{\text{pair}}\big) = \rho_g + \rho_d, \qquad r_1 + r_2 = 1 \qquad (6)$$

Here $r_1$ is the probability that the incoming atom is excited and can emit into the cavity, whereas $r_2$ is the probability that it is in the ground state and can absorb from the cavity. Their difference,

$$r_2 - r_1 = \frac{e^{\beta\hbar\omega} - e^{-\beta\hbar\omega}}{Z} = \frac{2\sinh(\beta\hbar\omega)}{Z}, \qquad (7)$$

sets the net damping bias imposed by the stream. Equation (7) shows explicitly that increasing the exchange coupling $|\lambda|$ increases the partition function $Z$, thereby reducing the effective population asymmetry at fixed $\omega$ and $T_{\text{atom}}$. In this one-atom coupling configuration, the off-diagonal pair coherence $\rho_{nd}$ does not enter the stream coefficients explicitly; its influence is indirect, through its contribution to the full pair state and through comparison with the two-atom coupling configuration considered later.

### 1.2 Detuning and the spectral-overlap filter

We now allow a finite detuning between the atomic transition and the cavity mode,

$$\Delta \equiv \omega - \omega_1 \qquad (8)$$

To derive the detuning dependence of the effective collision strength, it is useful to begin with the single-collision map itself. Over an interaction window of duration $\tau$, the joint atom–cavity evolution is generated by the Jaynes–Cummings interaction Hamiltonian of Eq. (4). The corresponding unitary is

$$U = \exp\left(-\frac{i}{\hbar} H_{\text{int}}\tau\right) \simeq I - \frac{i}{\hbar} H_{\text{int}}\tau - \frac{1}{2\hbar^2} H_{\text{int}}^2 \tau^2 + O(\phi^3), \qquad (9)$$

where $\phi = g\tau \ll 1$. For an incoming product state $\rho \otimes \rho_{\text{atom}}$, the reduced cavity state after one collision is

$$\Phi(\rho) = \text{Tr}_{\text{atom}}[U(\rho \otimes \rho_{\text{atom}})U^\dagger] \qquad (10)$$

To determine how finite detuning enters the effective collision strength, we now evaluate the leading nonvanishing contribution to the reduced map in Eq. (10), which arises at second order in the small parameter $\phi = g\tau$.

A useful observable for quantifying the effect of a single collision is the atom–cavity energy exchange, defined here as the change in cavity energy per collision,

$$\delta E \equiv \text{Tr}[H_c(\Phi(\rho) - \rho)] = \hbar\omega_1\, \text{Tr}[a^\dagger a(\Phi(\rho) - \rho)] \qquad (11)$$

This definition makes precise what is meant by "energy exchange" in the present collision model: $\delta E > 0$ corresponds to net energy delivered to the cavity during a single passage, while $\delta E < 0$ corresponds to net extraction.

The first-order term in the expansion of $U$ does not contribute to $\delta E$ or to the effective rate equation for the cavity occupation. The reason is that the incoming atomic pair carries no local dipole coherence on atom $A$: its reduced state is diagonal in the local energy basis, so that

$$\mathrm{Tr}_{\mathrm{atom}}(\sigma_A^{\pm}\rho_{\mathrm{atom}}) = 0.$$

Consequently, all first-order contributions are proportional to expectation values of the form $\langle\sigma_A^{\pm}\rangle$ and vanish after the partial trace. Physically, this means that there is no coherent drive term acting on the cavity from the incoming ancilla. The leading nonvanishing contribution therefore arises at second order, where emission and absorption processes appear through products such as $\sigma_A^{+}\sigma_A^{-}$ and $\sigma_A^{-}\sigma_A^{+}$, i.e., through the local excited- and ground-state weights $r_1$ and $r_2$ defined in Eq. (6).

To expose the detuning dependence, it is convenient to work in the interaction picture with respect to the uncoupled cavity and atom Hamiltonians. In that picture,

$$H_{\mathrm{int}}(t) = \hbar g(a\,\sigma_A^{+}e^{-i\Delta t} + a^{\dagger}\sigma_A^{-}e^{+i\Delta t}),$$

The second-order Dyson correction to the reduced collision map contains terms of the form

$$\int_0^{\tau} dt \int_0^{\tau} ds\ H_{\mathrm{int}}(t)\,\rho\,H_{\mathrm{int}}(s),$$

together with the corresponding anticommutator contributions. Multiplying the oscillatory factors from $H_{\mathrm{int}}(t)$ and $H_{\mathrm{int}}(s)$ generates the phase factor $e^{\pm i\Delta(t-s)}$. Because the cavity is damped by the ambient bath at rate $\kappa$, its correlation function decays as $e^{-(\kappa/2)|t-s|}$. As a result, the second-order correction to the single-collision map contains double integrals over the interaction window of the form

$$\int_0^{\tau} dt \int_0^{\tau} ds\ e^{-(\kappa/2)|t-s|}e^{\pm i\Delta(t-s)}. \qquad (12)$$

A compact derivation of this kernel from the second-order interaction-picture collision map is given in Appendix A.1. These kernels govern the detuning-dependent spectral overlap between the finite collision window and the cavity line.

Collecting these second-order contributions, one finds that the detuning dependence enters through the kernel

$$I(\Delta) = \int_0^{\tau} dt \int_0^{\tau} ds\ e^{-(\kappa/2)|t-s|}\,e^{i\Delta(t-s)} = 2\,\mathrm{Re}\int_0^{\tau}(\tau - u)\ e^{-(\kappa/2 - i\Delta)u}\,du, \qquad (13)$$

where $u = t - s$ has been introduced in the second form. The integral can be evaluated in closed form:

$$I(\Delta) = 2\,\mathrm{Re}\left\{\frac{\tau}{\alpha}(1 - e^{-\alpha\tau}) - \frac{1}{\alpha^2}[1 - e^{-\alpha\tau}(1 + \alpha\tau)]\right\}, \alpha \equiv \frac{\kappa}{2} - i\Delta \qquad (14)$$

Normalizing by $I(0)$, we define the spectral weight entering the collision rate as

$$L(\Delta) \equiv \frac{I(\Delta)}{I(0)}$$

The exact lineshape therefore interpolates smoothly between a cavity-limited Lorentzian and a time-window-limited $\mathrm{sinc}^2$ profile. In the coarse-grained Markov generator, only the slowly varying secular part is retained, and it is both standard and accurate to represent this interpolation by a single Lorentzian detuning filter,

$$L(\Delta) = \frac{1}{1 + (2\Delta/\Gamma_{\mathrm{over}})^2}, \Gamma_{\mathrm{over}} \equiv \kappa + \frac{1}{\tau} \qquad (15)$$

Here $\Gamma_{\mathrm{over}}$ is an effective overlap linewidth formed from the cavity homogeneous width $\kappa$ and the finite-time Fourier width $1/\tau$. This choice reproduces the correct limiting behavior: for $\tau \to \infty$, one recovers $L(\Delta) = 1/[1 + (2\Delta/\kappa)^2]$, while for $\kappa \to 0$ one obtains the expected time-window scaling $L(\Delta) \approx 1/[1 + (2\Delta\tau)^2]$. Any order-unity prefactors associated with different width conventions may be absorbed into $\Gamma_{\mathrm{over}}$ without changing the structure of the effective master equation.

Operationally, the detuning dresses the collision strength only through this overlap factor, so that in all stream-induced terms one makes the replacement

$$\phi^2 \longrightarrow \phi^2 L(\Delta) \qquad (16)$$

In other words, detuning does not alter the form of the collision model; it simply rescales the effective pump and loss generated by each ancilla passage according to the spectral overlap between the atomic transition and the damped cavity mode.

Further details on the detuning-overlap kernel, including its cavity-limited and time-window-limited asymptotic forms, as well as additional consistency checks for the one-atom coupling configuration, are provided in Appendix A.

### 1.3 Collision model and effective Lindblad generator

Using the detuning-dressed replacement introduced in Eq. (16), the single-passage map obtained from the unitary expansion and partial trace can now be written in compact form. Retaining terms up to $O(\phi^2)$, the reduced cavity state after one collision is

$$\Phi(\rho) \simeq \rho + \phi^2 L(\Delta)[r_2\, \mathcal{D}[a]\rho + r_1\, \mathcal{D}[a^\dagger]\rho], \qquad (17)$$

where

$$\mathcal{D}[L]\rho \equiv L\rho L^\dagger - \frac{1}{2}\{L^\dagger L, \rho\}$$

is the usual Lindblad dissipator. Equation (17) shows that, within the weak-collision approximation, the sole effect of detuning is to renormalize the effective collision strength by the overlap factor $L(\Delta)$, while the operator structure of the collision map remains unchanged. The coefficient $r_1$ multiplies the excitation channel $\mathcal{D}[a^\dagger]$, corresponding to photon injection into the cavity by an excited incoming atom, whereas $r_2$ multiplies the loss channel $\mathcal{D}[a]$, corresponding to photon absorption by an atom initially in its ground state.

When independently prepared atom pairs arrive as a Poisson process with rate $R$, the corresponding coarse-grained stream generator is obtained in the standard collision-model limit by multiplying the one-passage map by the arrival rate. The stream therefore contributes[13]

$$\mathcal{L}_{\text{stream}}\rho = R\, \phi^2 L(\Delta)[r_2\, \mathcal{D}[a]\rho + r_1\, \mathcal{D}[a^\dagger]\rho]. \qquad (18)$$

This is the effective Lindblad description of the engineered atomic reservoir in the one-atom coupling configuration.

In parallel, the cavity is continuously coupled to the ambient phonon bath at temperature $T_{\text{bath}}$. As discussed above, this thermal environment is represented by

$$\mathcal{L}_{\text{bath}}\rho = \kappa^{(-)}\mathcal{D}[a]\rho + \kappa^{(+)}\mathcal{D}[a^\dagger]\rho, \kappa^{(-)} = \kappa(\bar{n}_1 + 1), \kappa^{(+)} = \kappa\bar{n}_1, \qquad (19)$$

which satisfies the detailed-balance relation $\kappa^{(+)}/\kappa^{(-)} = \exp\,[-\hbar\omega_1/(k_B T_{\text{bath}})]$. The full reduced master equation for the cavity density operator is therefore

$$\dot{\rho} = \mathcal{L}_{\text{bath}}\rho + \mathcal{L}_{\text{stream}}\rho. \qquad (20)$$

### 1.4 Photon-number dynamics and cavity effective temperature

We now characterize the cavity state by its photon number

$$n \equiv \langle a^\dagger a\rangle = \text{Tr}(a^\dagger a\, \rho).$$

Using the standard identities

$$\text{Tr}[a^\dagger a\, \mathcal{D}[a]\rho] = -n, \text{Tr}[a^\dagger a\, \mathcal{D}[a^\dagger]\rho] = n + 1,$$

which follow from $aa^\dagger = a^\dagger a + 1$, Eq. (20) closes exactly on $n$. The resulting rate equation is

$$\frac{dn}{dt} = -\kappa(n - \bar{n}_1) + Rr_1\phi^2 L(\Delta) - R(r_2 - r_1)\phi^2 L(\Delta)\, n. \qquad (21)$$

This equation has a transparent physical interpretation. The first term is the relaxation of the cavity toward the thermal occupation $\bar{n}_1$ imposed by the phonon bath. The second term is an incoherent pump from the atom stream, proportional to the excited-state weight $r_1$. The third term is the stream-induced damping, whose magnitude is controlled by the imbalance $r_2 - r_1$. Detuning enters only through the common overlap factor $L(\Delta)$, reducing both the pump and damping contributions of the stream without changing their relative bias.

It is useful to group the coefficients into an effective damping and an effective injection rate,

$$\Gamma_{\downarrow}(\Delta) \equiv \kappa + R(r_2 - r_1)\phi^2 L(\Delta), J_{\uparrow}(\Delta) \equiv \kappa \bar{n}_1 + R r_1 \phi^2 L(\Delta), \qquad (22)$$

so that the photon-number equation takes the compact birth–death form

$$\frac{dn}{dt} = -\Gamma_{\downarrow}(\Delta)\, n + J_{\uparrow}(\Delta). \qquad (23)$$

For an arbitrary initial condition $n(0)$, the solution is

$$n(t) = n^*(\Delta) + [n(0) - n^*(\Delta)]e^{-\Gamma_{\downarrow}(\Delta)t}, \qquad n^*(\Delta) = \frac{J_{\uparrow}(\Delta)}{\Gamma_{\downarrow}(\Delta)} = \frac{\kappa \bar{n}_1 + R r_1 \phi^2 L(\Delta)}{\kappa + R(r_2 - r_1)\phi^2 L(\Delta)} \qquad (24)$$

Thus the cavity relaxes exponentially toward a unique steady state provided $\Gamma_{\downarrow}(\Delta) > 0$. For noninverted incoming atoms, $r_2 \geq r_1$, this condition is automatically satisfied. If the stream is inverted, $r_1 > r_2$, stability requires that the intrinsic cavity damping $\kappa$ exceed the net negative damping induced by the stream.

The steady state may be parametrized by an effective Bose–Einstein temperature at the cavity frequency $\omega_1$. Writing

$$n^*(\Delta) = \frac{1}{e^{\frac{\hbar\omega_1}{[k_B T_{\text{cav}}(\Delta)]}} - 1}, \qquad (25)$$

one obtains

$$T_{\text{cav}}(\Delta) = \frac{\hbar\omega_1}{k_B \ln\left(1 + 1/n^*(\Delta)\right)} \qquad (26)$$

Eqs. (25) and (26) provide the central operational quantity for the one-atom coupling configuration: they give the effective cavity temperature resulting from the competition between the thermal phonon bath and the detuning-dressed, correlation-influenced stream of ancilla pairs.

### 1.5 Physical interpretation

Equation (21) shows that the one-atom stream acts as an incoherent reservoir for the cavity. The term $R r_1 \phi^2 L(\Delta)$ is an effective pump generated by excited incoming atoms through the channel $\mathcal{D}[a^\dagger]$, whereas $R r_2 \phi^2 L(\Delta)$ is an additional loss channel generated by ground-state atoms through $\mathcal{D}[a]$. Their difference, $R(r_2 - r_1)\phi^2 L(\Delta)$, determines whether the stream augments the cavity damping ($r_2 > r_1$, cooling tendency) or acts as net gain ($r_1 > r_2$, heating tendency). The ambient phonon bath fixes the baseline occupation $\bar{n}_1$ and contributes the additive damping $\kappa$, which stabilizes the dynamics provided $\kappa > R(r_1 - r_2)\phi^2 L(\Delta)$ for an inverted stream. The detuning factor $L(\Delta)$ gives a direct measure of spectral overlap: only the fraction of the collision strength that overlaps the damped cavity line contributes effectively to pumping and loss.

### 1.6 Validity of the detuned description

The above detuned collision model relies on several standard approximations. First, the interaction is assumed weak and short, $\phi = g\tau \ll 1$, so that the single-collision map may be truncated at second order. Second, the ancilla pairs arrive independently with $R\tau \ll 1$, allowing a Markov coarse graining with finite $R\phi^2$. Third, the atom and cavity are assumed near resonance on the scale of the rotating-wave approximation, $|\omega - \omega_1| \ll \omega_1$. Fourth, a Born approximation is used, so cavity–stream correlations do not accumulate between successive collisions. Within this regime, Eqs. (21)–(26) provide a controlled detuned generalization of the resonant one-atom result. The effective overlap width $\Gamma_{\text{over}} = \kappa + 1/\tau$ should be understood as a coarse-grained phenomenological interpolation between cavity-limited and time-window-limited broadening; alternative width conventions differ only by order-unity factors and do not modify the structure of the resulting master equation.

**2. Correlated atom-pair stream: both atoms coupled to a phonon-tethered cavity (with detuning)**

We now consider the second interaction geometry, in which both atoms of each correlated pair couple to the same cavity mode during the collision window. As in Section 1, the cavity is described by a single bosonic mode $a$ of frequency $\omega_1$, continuously coupled to the ambient phonon bath at temperature $T_{\mathrm{bath}}$ with bath occupation $\bar{n}_1$ given by Eq. (1). The incoming pair state is the same thermal correlated state introduced in Eqs. (2)-(3), with matrix elements $\rho_e$, $\rho_g$, $\rho_d$, and $\rho_{nd}$.

If the two atoms couple with equal strength $g$ to the same cavity mode, the interaction Hamiltonian becomes

$$H_{\mathrm{int}}^{(2)} = \hbar g[a(\sigma_A^+ + \sigma_B^+) + a^\dagger(\sigma_A^- + \sigma_B^-)] = \hbar g(aS_+ + a^\dagger S_-), \qquad (27)$$

where

$$S_\pm \equiv \sigma_A^\pm + \sigma_B^\pm$$

are the collective ladder operators of the pair. No additional operator of the form $\sigma_A^\pm + \sigma_B^\mp$ is required: under the Jaynes–Cummings rotating-wave approximation, the cavity couples only to the co-rotating collective dipole $S_\pm$. The exchange-type cross terms $\sigma_A^+\sigma_B^-$ and $\sigma_B^+\sigma_A^-$ already appear explicitly when forming the correlators $S_+S_-$ and $S_-S_+$ used below to define the effective two-subsystem stream coefficients.

As in the one-atom case, we work in the weak, short-collision limit $\phi = g\tau \ll 1$. When both atoms couple during the same interaction window, the net second-order kick on the cavity depends on their relative geometry and phase. We parametrize this by an effective two-atom collision strength

$$\phi_2^2 \equiv \chi\, \phi^2, 0 \le \chi \le 2, \qquad (28)$$

where $\chi$ is a phenomenological interference factor. The value $\chi = 2$ corresponds to equal, in-phase couplings that add constructively, $\chi = 1$ corresponds to effectively independent addition, and $0 < \chi < 1$ describes partial cancellation due to geometry, phase mismatch, or timing offsets. In the numerical results presented here, unless stated otherwise, we use the baseline choice $\chi = 1$.

The incoming pairs are again assumed to arrive as a Poisson process with rate $R$.

The effective two-atom stream coefficients are defined as the normalized collective emission and absorption weights of the incoming pair,

$$r_1^{(2)} \equiv \frac{1}{2}\,\mathrm{Tr}\big(S_+S_-\,\rho_{\mathrm{pair}}\big), \qquad r_2^{(2)} \equiv \frac{1}{2}\,\mathrm{Tr}\big(S_-S_+\,\rho_{\mathrm{pair}}\big), \qquad (29)$$

which, for the thermal pair state of Eqs. (2)–(3), evaluate to

$$r_1^{(2)} = \rho_e + \rho_d + \rho_{nd}, \qquad r_2^{(2)} = \rho_g + \rho_d + \rho_{nd}. \qquad (30)$$

The factor $1/2$ in Eq. (29) is a convenient normalization that keeps the two-atom coefficients on the same footing as the one-atom coefficients $r_1$ and $r_2$. Two useful identities follow immediately:

$$r_2^{(2)} - r_1^{(2)} = \rho_g - \rho_e = \frac{2\sinh(\beta\hbar\omega)}{Z} > 0 \qquad (\beta\hbar\omega > 0), \qquad (31)$$

$$r_1^{(2)} + r_2^{(2)} = 1 + 2\rho_{nd}. \qquad (32)$$

Equation (31) shows that the net damping bias is still controlled only by the population asymmetry of the pair, whereas Eq. (32) shows that the absolute magnitude of the stream-induced pump and loss is shifted by the inter-atomic coherence $\rho_{nd}$.

As in Section 1, finite detuning enters through the same overlap filter $L(\Delta)$, with $\Delta = \omega - \omega_1$ and $L(\Delta)$ given by Eq. (15). The detuning therefore dresses the two-atom collision strength by the replacement $\phi_2^2 \to \phi_2^2 L(\Delta)$. Expanding the one-passage unitary generated by Eq. (27) to second order in $\phi$ and tracing over the pair yields the reduced cavity map

$$\Phi^{(2)}(\rho) \simeq \rho + \phi_2^2 L(\Delta)\left[r_2^{(2)}\mathcal{D}[a]\rho + r_1^{(2)}\mathcal{D}[a^\dagger]\rho\right]. \qquad (33)$$

Thus, just as in the one-atom configuration, the collective stream contributes an additional Lindblad pump–loss pair, but now with coefficients that depend explicitly on the pair coherence $\rho_{nd}$.

For a Poisson train of collisions with rate $R$, the corresponding coarse-grained stream generator is

$$\mathcal{L}_{\text{stream}}^{(2)}\rho = R\phi_2^2 L(\Delta)\left[r_2^{(2)}\mathcal{D}[a]\rho + r_1^{(2)}\mathcal{D}[a^\dagger]\rho\right]. \qquad (34)$$

The ambient phonon-bath contribution remains exactly as in Section 1,

$$\mathcal{L}_{\text{bath}}\rho = \kappa^{(-)}\mathcal{D}[a]\rho + \kappa^{(+)}\mathcal{D}[a^\dagger]\rho, \kappa^{(-)} = \kappa(\bar{n}_1 + 1), \kappa^{(+)} = \kappa\bar{n}_1, \qquad (35)$$

so that the full reduced master equation is

$$\dot{\rho} = \mathcal{L}_{\text{bath}}\rho + \mathcal{L}_{\text{stream}}^{(2)}\rho. \qquad (36)$$

## 2.1 Photon-number dynamics and cavity effective temperature

Defining the cavity photon number as before,

$$n \equiv \langle a^\dagger a\rangle,$$

and using the same adjoint identities employed in Section 1, Eq. (36) closes on $n$ as

$$\frac{dn}{dt} = -\kappa(n - \bar{n}_1) + Rr_1^{(2)}\phi_2^2 L(\Delta) - R(r_2^{(2)} - r_1^{(2)})\phi_2^2 L(\Delta)\, n. \qquad (37)$$

This is the exact two-atom analogue of Eq. (21), now derived explicitly rather than obtained by substitution. It is useful to define the detuning-dressed damping and injection rates

$$\Gamma_\downarrow^{(2)}(\Delta) \equiv \kappa + R(r_2^{(2)} - r_1^{(2)})\phi_2^2 L(\Delta), \qquad (38)$$

$$J_\uparrow^{(2)}(\Delta) \equiv \kappa\bar{n}_1 + Rr_1^{(2)}\phi_2^2 L(\Delta), \qquad (39)$$

so that the photon-number dynamics take the compact birth–death form

$$\frac{dn}{dt} = -\Gamma_\downarrow^{(2)}(\Delta)\, n + J_\uparrow^{(2)}(\Delta). \qquad (40)$$

The solution for an arbitrary initial condition $n(0)$is therefore

$$n(t) = n^*(\Delta) + [n(0) - n^*(\Delta)]e^{-\Gamma_\downarrow^{(2)}(\Delta)t}, \qquad (41)$$

with steady-state occupation

$$n^*(\Delta) = \frac{J_\uparrow^{(2)}(\Delta)}{\Gamma_\downarrow^{(2)}(\Delta)} = \frac{\kappa\bar{n}_1 + Rr_1^{(2)}\phi_2^2 L(\Delta)}{\kappa + R(r_2^{(2)} - r_1^{(2)})\phi_2^2 L(\Delta)}. \qquad (42)$$

As in Section 1, the cavity effective temperature is defined by fitting this steady state to a Bose–Einstein form at frequency $\omega_1$,

$$T_{\text{cav}}(\Delta) = \frac{\hbar\omega_1}{k_B \ln(1 + 1/n^*(\Delta))}. \qquad (43)$$

## 2.2 Physical role of $\rho_{nd}$ and $\chi$

The distinctive feature of the two-atom configuration is that the inter-atomic coherence $\rho_{nd}$ enters the effective stream coefficients directly through Eq. (30). Because it appears with the same sign in both $r_1^{(2)}$and $r_2^{(2)}$, it does not change the net bias $r_2^{(2)} - r_1^{(2)}$, which remains fixed by the population asymmetry of the pair, but it does change the absolute magnitude of the stream-induced pump and loss. In this sense, $\rho_{nd}$ controls how strongly the correlated pair engages the cavity without changing whether the stream tends to cool or heat it. For $\rho_{nd} < 0$, both $r_1^{(2)}$ and $r_2^{(2)}$ are lowered relative to the one-atom case, shifting the detailed balance seen by the cavity toward colder steady states when the stream dominates over the phonon bath.

The parameter $\chi$ plays a distinct role. It contains only geometry and phase information and sets the overall scale of the collective second-order kick. Thus $\rho_{nd}$ is a thermodynamic property of the incoming pair state, while $\chi$ is an engineering parameter of the interaction region. The detuning factor $L(\Delta)$ multiplies both on the same footing, acting as a spectroscopic gate that uniformly suppresses the stream-induced pump and damping away from resonance. The three quantities $\rho_{nd}$, $\chi$, and $L(\Delta)$ therefore control, respectively, the pair-state coherence, the collective coupling enhancement, and the spectral overlap with the cavity.

### 2.3 Validity of the detuned two-atom description

The above derivation relies on the same approximations as Section 1: weak, short collisions $\phi = g\tau \ll 1$; Poissonian arrivals with $R\tau \ll 1$ and finite $R\phi_2^2$ enabling Markov coarse graining; near-resonant coupling consistent with the rotating-wave approximation; and a Born treatment in which cavity–stream correlations do not accumulate between collisions. Within this regime, Eqs. (37)–(43) provide the controlled two-atom counterpart of the one-atom result, now with explicit dependence on pair coherence through $r_1^{(2)}$ and $r_2^{(2)}$.

Additional limiting cases and a step-by-step evaluation workflow for the two-atom steady state are given in Appendix B.

## 3. Results and Discussions

The numerical results presented below are obtained by direct evaluation of the closed-form analytic expressions derived in Secs. 1 and 2; additional limiting cases and practical evaluation details for the two-atom configuration are provided in Appendix B.

To ensure consistency across parameter sweeps, a common baseline parameter set is adopted and summarized in Table 1. These values, representative of contemporary circuit-QED devices, place the system firmly within the weak-collision regime ($\phi = g\tau \ll 1$) and provide a realistic reference point for assessing cooling performance as individual parameters (detuning, cavity–bath coupling, atom–cavity strength, and pair correlations) are varied.

Table 1: Global Baseline Parameters Used in Analytical Evaluations

| Quantity | Symbol / Definition | Value | Notes |
|---|---|---|---|
| Bath temperature | $T_{\mathrm{bath}}$ | 1 K | Cryogenic stage (phonon environment) |
| Atomic (two-qubit) temperature | $T_{\mathrm{atom}}$ | 50 mK | Effective temperature of the engineered reservoir |
| Cavity frequency | $\omega_1/2\pi$ | 5 GHz | Fundamental cavity mode |
| Atom–cavity interaction time | $\tau$ | 50 ns | Set by flux-pulse duration |
| Collision (arrival) rate | $R$ | 5×10$^6$ 1/s | Mean pair-arrival frequency |
| Inter-qubit coupling | $\lambda$ | 5 GHz | Exchange coupling between qubits |
| Jaynes–Cummings coupling | $g/2\pi$ | 0.5 MHz | Coherent qubit–cavity interaction strength |
| Cavity–bath damping rate | $\kappa/2\pi$ | 100 Hz | Moderately weak phonon tether |
| Collision angle | $\phi = g\tau$ | $\phi \approx 0.16$ | Ensures weak-collision limit ($\phi \ll 1$) |
| Two-atom enhancement factor | $\chi$ | 1 | Baseline calculations use $\phi_2^2 = \chi\phi^2 = \phi^2$ |

### 3.1 Cooling-heating crossover induced by atom-cavity detuning

Figure 2 (a) and (b) summarize the central mechanism by which a correlated atom stream modifies the steady-state temperature of a phonon-tethered cavity. The key control parameter is the detuning $\Delta = \omega - \omega_1$, which governs both the efficiency and directionality of energy exchange between the incoming atomic state and the cavity mode. Because the single-interaction rotation angle $\phi = g\tau \ll 1$ places the system firmly in the weak-collision regime, the effect of each atomic pair on the cavity is additive and can be described by an effective birth–death process for the cavity photon number $n^*(\Delta)$.

As shown in Figure 2 (a), the ratio $T_{\mathrm{cav}}/T_{\mathrm{atom}}$ for the one-atom interaction model exhibits a shallow minimum near $\Delta \approx 0$. This behavior is captured analytically by Eq. (24), with the detuning dependence entering through the Lorentzian spectral-overlap filter L(Δ). Near resonance, the cavity efficiently absorbs entropy-carrying excitations from the incoming atom pairs, and the stream-induced detailed balance

$$\frac{A_\uparrow}{A_\downarrow - A_\uparrow} = \frac{n^*}{1+n^*}$$

with $A_\uparrow(\Delta) = \kappa\bar{n}_1 + Rr_1\phi^2 L(\Delta)$, $A_\downarrow(\Delta) = \kappa(\bar{n}_1 + 1) + Rr_2\phi^2 L(\Delta)$, becomes dominant over the intrinsic phonon-bath contribution. As a result, the cavity is pulled toward the effective temperature imposed by the correlated stream, lowering $T_{\text{cav}}$ significantly below the phonon-bath temperature.

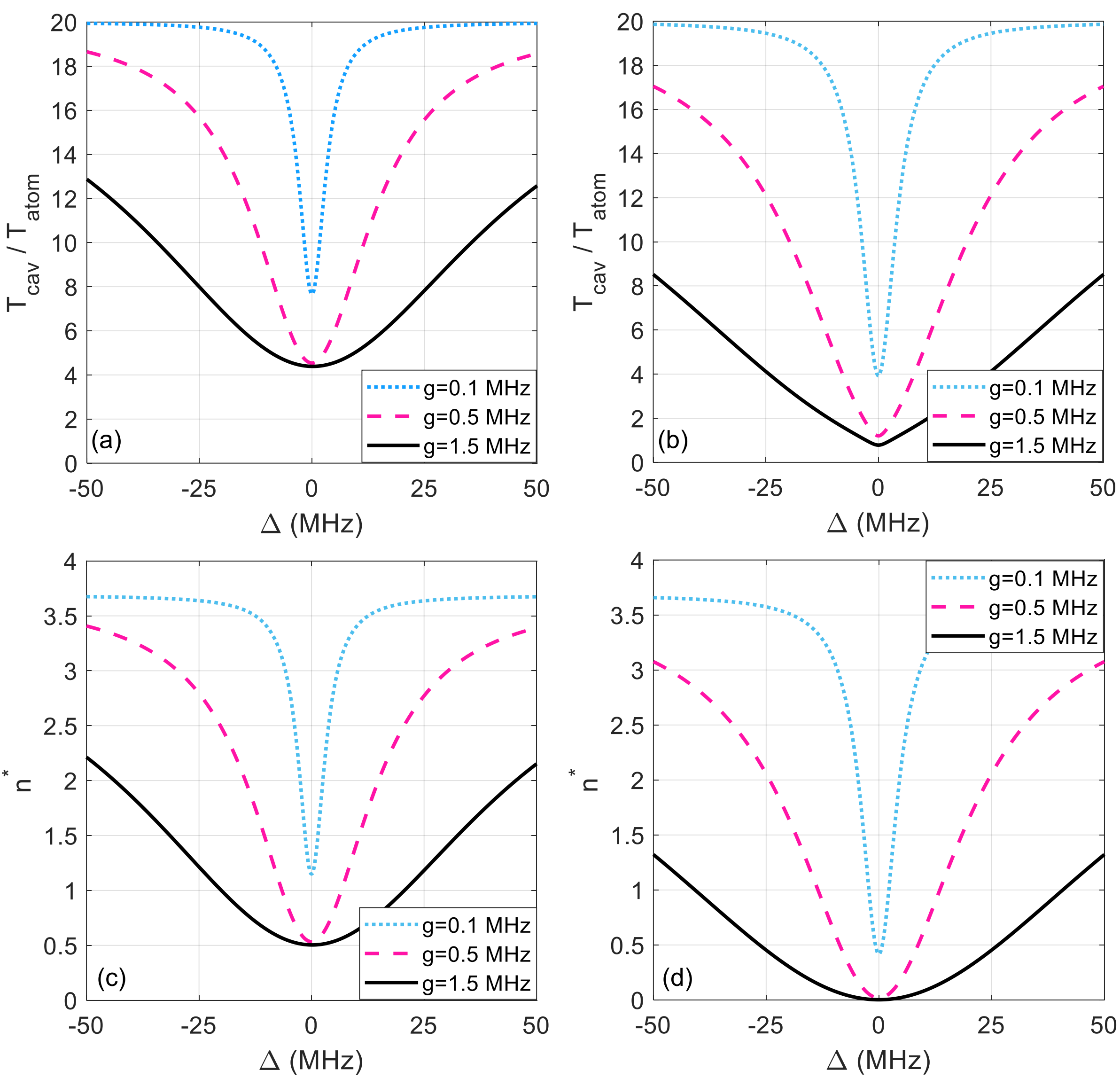

Figure 2: (a,b) Effective cavity temperature versus atom-cavity detuning. The ratio $\mathrm{T_{cav}/T_{atom}}$ is shown as a function of the detuning $\Delta = \omega - \omega_1$ for one-atom (a) and two-atom (b) interaction models. Near resonance ($\Delta = 0$), the correlated atom-pair reservoir efficiently exchanges energy with the cavity and drives it toward the reservoir's temperature. For large detuning, the coherent exchange channel is suppressed, the phonon bath dominates, and the cavity warms, yielding $\mathrm{T_{cav}} > \mathrm{T_{atom}}$. Three representative coupling strengths $\mathrm{g}$ are shown in each panel, demonstrating stronger cooling for larger $g$. In the one-atom case (a), the minimum $\mathrm{T_{cav}/T_{atom}}$ remains above unity, approaching a value near $\sim 4$. In contrast, in the two-atom case (b), $\mathrm{T_{cav}/T_{atom}}$ falls below unity, reaching values near $\sim 0.5$ near resonance and for stronger coupling, indicating genuine refrigeration of the cavity below the atom temperature. (c,d) Steady-state cavity photon number versus detuning. The steady-state cavity occupation $n^*$ is plotted for the same parameters as in (a) and (b), for the one-atom (c) and two-atom (d) models respectively. The detuning dependence reflects the spectral overlap between the atom and cavity modes: efficient near resonance, resulting in reduced photon populations, and suppressed at large detuning, where the cavity relaxes toward the phonon bath. The behavior of $n^*(\Delta)$ corresponds directly to the cooling and heating trends observed in panels (a) and (b).

For detunings $|\Delta| \gtrsim \Gamma_{\text{over}}$, the stream–cavity energy exchange ($\delta E$) is strongly suppressed and $L(\Delta) \to 0$, such that the steady state population $n^*$ reduces to the phonon-tethered value $n^* \to \bar{n}_1$. This produces the symmetric heating wings observed in Figure 2 (a): because the stream no longer removes sufficient entropy to compensate bath-induced excitations, the cavity relaxes toward the hotter phonon bath. Thus, the cooling-heating profile as a function of detuning is an intrinsic consequence of spectral filtering imposed by the weak-collision channel.

The influence of the coupling strength $g$ is highlighted in Figure 2 (a). Increasing $g$ enhances the per-collision rotation $g\tau$, thereby amplifying the stream-induced Lindblad rates $R\phi^2 r_1$ and $R\phi^2(r_2 - r_1)$. Physically, this enlarges the domain in which the atomic stream overcomes phonon-bath heating. As $g$ increases, the minimum of $T_{\text{cav}}/T_{\text{atom}}$ becomes increasingly pronounced, and efficient cooling persists out to larger values of detuning. This behavior is consistent with the analytic scaling in Eqs. (24)-(26), which interpolates between a bath-dominated steady state when $g\tau \to 0$ and a stream-dominated regime when $g\tau$ is sufficiently large for the engineered reservoir to dictate the detailed balance.

Figure 2 (b) also reveals the qualitative enhancement provided by the two-atom interaction model, where both members of the correlated pair couple to the cavity. Here the effective upward and downward rates depend on the two-subsystem coefficients $\left(r_1^{(2)}, r_2^{(2)}\right)$, which differ from their one-atom counterparts because both atoms participate in the cavity interaction. As a result, the cavity can be cooled below the temperature of the atomic reservoir itself, with $T_{\text{cav}}/T_{\text{atom}}$ approaching values as low as $\sim 0.5$ near resonance for strong coupling. The cooling region is also broader: the two-qubit reservoir maintains dominance over the phonon bath across a wider detuning interval, reflecting its stronger collective interaction with the cavity.

Figure 2 (c) and (d) plot the corresponding steady-state photon number $n^*(\Delta)$ for the one-atom and two-atom cases, respectively. The photon-number trends mirror the temperature behavior: in Figure 2 (c), the minimum of $n^*$is modest and remains above the stream-imposed value, whereas in Figure 2 (d) the cavity photon population is strongly suppressed near resonance, consistent with the significantly deeper cooling achieved in the two-atom configuration. The close correspondence between $n^*(\Delta)$ and $T_{\text{cav}}(\Delta)$ across panels (a)–(d) is expected, since $T_{\text{cav}}$ is obtained directly from $n^*$ by Bose inversion. Within the present model, the stronger refrigeration in the two-atom case is therefore traced to the correlation-modified coefficients $r_1^{(2)}$and $r_2^{(2)}$, rather than to any change in the structure of the dissipative dynamics.

Together, Figure 2 (a–d) reveal the essential physics of both the one-subsystem and two-subsystem models. In the one-atom case, panels (a) and (c) show that the cavity temperature is governed by a competition between (i) resonant entropy extraction by the correlated atomic stream and (ii) thermal injection from the phonon bath. Detuning suppresses the former through the Lorentzian spectral filter $L(\Delta)$, while increasing coupling $g$ strengthens it by enhancing the stream-induced Lindblad rates $R\phi^2 r_1$ and $R\phi^2(r_2 - r_1)$. As a result, the cooling minimum at $\Delta = 0$ marks the operational "sweet spot'' of the refrigerator, where the stream imposes a nonequilibrium detailed balance that can pull the cavity temperature significantly below that of the phonon reservoir, though still above the atom-stream temperature $T_{\text{atom}}$.

In contrast, panels (b) and (d) highlight the two-atom interaction case, where both constituents of the correlated pair couple to the cavity. The collective interaction modifies the effective upward–downward rates via the two-subsystem coefficients $\left(r_1^{(2)}, r_2^{(2)}\right)$, enabling a substantially deeper cooling window. At resonance, the engineered reservoir is strong enough to impose a detailed balance colder than the atomic reservoir itself, allowing $T_{\text{cav}}$ to drop below $T_{\text{atom}}$ with minima near $T_{\text{cav}}/T_{\text{atom}} \sim 0.5$. This enhanced cooling capability reflects the modified effective transition rates of the collective two-atom coupling configuration, which broaden the detuning range over which the stream dominates over the phonon bath. The specific role of intra-pair coupling and coherence is analyzed in Sec. 3.3. Consequently, while the one-atom model demonstrates stream-assisted refrigeration of a phonon-tethered cavity, the two-atom model reveals a qualitatively stronger regime in which correlations within each incoming atomic pair enable genuine quantum-enhanced refrigeration.

### 3.2 Competition between phonon heating and stream-induced cooling

Figure 3 (a–d) illustrates how the cavity steady state emerges from a competition between two distinct energy-exchange channels: (i) thermalization with the environmental phonon bath, and (ii) entropy extraction by the correlated atomic reservoir. The relative strength of these channels determines whether the cavity is ultimately pulled toward the hot phonon bath or toward the colder effective temperature imposed by the engineered atom stream.

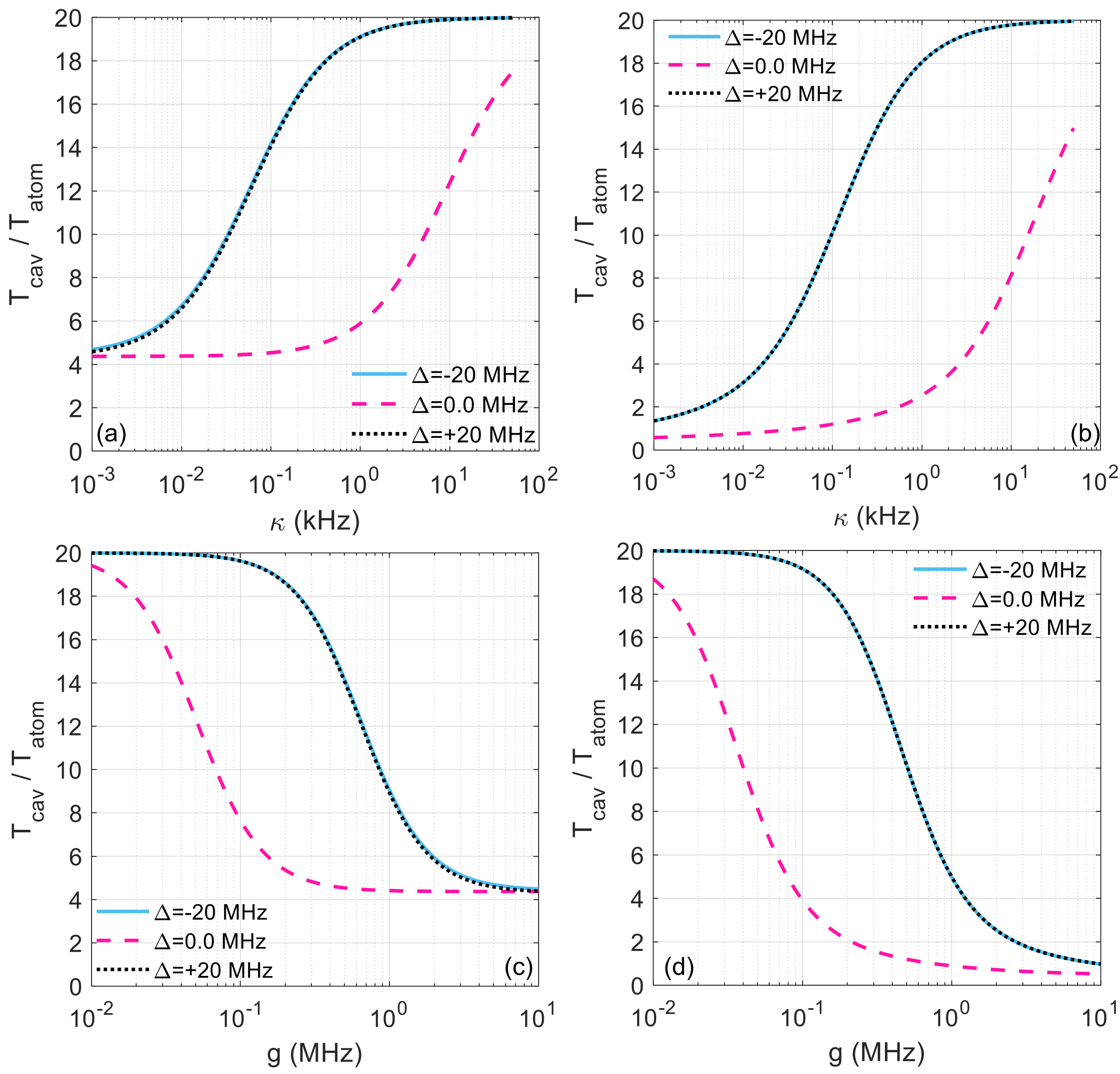

Figure 3: *(a,b) Dependence of cavity cooling on phonon coupling.* The steady-state temperature ratio $T_{\rm cav}/T_{\rm atom}$ is shown as a function of the cavity–phonon energy-damping rate $\kappa$ for the one-atom (a) and two-atom (b) interaction models. In the weak-phonon-coupling regime, the correlated atomic stream dominates the energy balance and drives the cavity far from the environmental phonon bath. In the one-atom case (a), $T_{\rm cav}/T_{\rm atom}$ saturates near $\sim 4$ as $\kappa \to 0$, while in the two-atom case (b) the collective interaction enables genuine refrigeration, with $T_{\rm cav}/T_{\rm atom} \to$

0.5 in the same limit. As $\kappa$ increases, the phonon bath increasingly overwhelms the stream-induced cooling and forces the cavity back toward its thermal equilibrium value $T_{\mathrm{cav}} \to T_{\mathrm{bath}}$. The crossover marks the departure from the idealized $\kappa = 0$ limit and quantifies the practical dissipation pathways present in superconducting microwave cavities. *(c,d) Dependence on cavity-atom coupling strength $g$.* Panels (c) and (d) show $T_{\mathrm{cav}}/T_{\mathrm{atom}}$ versus coupling $g$ for the one-atom and two-atom models, respectively. Increasing $g$ strengthens each weak collision $\phi = g\tau \ll 1$, amplifying the stream-induced Lindblad rates and thereby deepening the cavity cooling window. In both cases the curves saturate at the same limiting values observed in (a) and (b), near $\sim 4$ for the one-atom model and $\sim 0.5$ for the two-atom model, reflecting the intrinsic balance between excitation absorption and emission imposed by the correlated atomic reservoir. In the weak-coupling regime the stream cannot compete with phonon heating, while stronger $g$ enables the engineered reservoir to impose its own nonequilibrium detailed balance on the cavity mode. These panels highlight an experimentally tunable transition between bath-dominated and reservoir-dominated cavity steady states.

In the one-atom model, Figure 3 (a) shows the dependence of $T_{\mathrm{cav}}/T_{\mathrm{atom}}$ on the cavity–phonon damping rate $\kappa$. For very small $\kappa$, the cavity scarcely exchanges energy with the phonons, and the correlated atomic stream fully dominates the energy balance. In this limit the cavity saturates at the intrinsic stream-imposed detailed balance, yielding $T_{\mathrm{cav}}/T_{\mathrm{atom}} \to 4$. This reproduces the behavior predicted in earlier idealized analyses that neglected external baths.[19] As $\kappa$ increases, however, the phonon bath injects excitations that gradually overwhelm the weak-collision channel, forcing the cavity back toward the hotter environmental temperature. The smooth crossover in Figure 3 (a) therefore captures the realistic dissipation pathways present in superconducting microwave resonators, where phonon leakage cannot be ignored.

The two-atom model in Figure 3 (b) exhibits the same qualitative trend but with a dramatically different limiting behavior. When $\kappa \to 0$, the cavity is no longer driven toward a hotter steady state, but instead is cooled below the atom-pair temperature, reaching $T_{\mathrm{cav}}/T_{\mathrm{atom}} \to 0.5$. This quantum-enhanced refrigeration arises from the collective interaction of both atoms with the cavity, which modifies the upward-downward transition rates and allows the correlated pair to impose a colder effective detailed balance on the mode. As $\kappa$ increases, phonon-induced excitations again dominate, eventually washing out the cooling advantage and restoring $T_{\mathrm{cav}} \to T_{\mathrm{bath}}$.

Panels (c) and (d) examine the complementary dependence on the cavity-atom coupling strength $g$. Increasing $g$ enhances the per-collision rotation angle $\phi = g\tau$, amplifying the stream-induced Lindblad rates that govern the cooling dynamics. In the one-atom model (c), weak coupling leaves the cavity close to the phonon-dominated state, while stronger coupling drives it toward the saturated limit near $T_{\mathrm{cav}}/T_{\mathrm{atom}} \sim 4$. In the two-atom case (d), the same trend is observed, but the saturation occurs at the much colder value $T_{\mathrm{cav}}/T_{\mathrm{atom}} \sim 0.5$. Increasing $g$ therefore extends the domain where the engineered reservoir dominates over the phonon bath and reveals the stronger cooling power of the collective two-atom interaction.

Taken together, Figure 3 (a–d) identifies the two essential experimental control knobs for nonequilibrium cavity refrigeration: (i) the suppression of unwanted phonon coupling ($\kappa$), and (ii) the enhancement of intentional atom-cavity interactions ($g$). Efficient cooling requires the engineered reservoir to dominate over environmental heating. This can be achieved either by operating with high cavity quality factors (small $\kappa$) or by increasing $g$ so that the correlated atomic stream imposes its own detailed balance. The two-atom model demonstrates that collective correlations can significantly outperform the one-atom case, enabling the cavity temperature to be driven below the atomic reservoir temperature, an important capability for integrating such cavities with thermally sensitive quantum devices such as quantum dots, spin qubits, and bosonic memories.

### 3.3 Weak influence of pair correlations and global structure of the cooling landscape

Figure 4 (a–d) provides the explicit parameter study that substantiates the role of intra-pair coupling and coherence in the cooling mechanism. Whereas detuning and phonon damping govern the dominant cooling-heating balance, the intra-pair exchange interaction $\lambda$ modulates the internal structure of the correlated atomic reservoir and therefore affects the detailed balance imposed on the cavity. The effect, however, depends critically on whether one or both atoms interact with the cavity.

In the one-atom model, Figure 4 (a) shows that $T_{\mathrm{cav}}/T_{\mathrm{atom}}$ exhibits only a weak dependence on $\lambda$, and in fact increases slightly as $\lambda$ grows. This behavior follows from the structure of the arrival coefficients in the one-subsystem configuration. Although the incoming thermal pair is described by the weights $(\rho_e, \rho_g, \rho_d, \rho_{nd})$ of Eqs. (2)-(3), only one spin interacts with the cavity, so the effective upward and downward weights reduce to $r_1 = \rho_e + \rho_d$ and $r_2 = \rho_g + \rho_d$. The coherence term $\rho_{nd}$, responsible for the correlated advantages in the two-atom case, drops out entirely. Consequently, the cooling-relevant difference $r_2 - r_1 = \rho_g - \rho_e$ depends only on the single-spin splitting $\omega$ and is independent of $\lambda$ except for normalization via the partition function $Z$. Increasing $\lambda$ increases the weight of the $\rho_d$ manifold, which appears symmetrically in both $r_1$ and $r_2$. The net effect is a mild reduction in stream-induced cooling efficiency, hence the slow rise of $T_{\text{cav}}/T_{\text{atom}}$ with increasing $\lambda$.

The situation is qualitatively different in the two-atom model, where both atoms interact simultaneously with the cavity. In this case (Figure 4 (b)), the stream-induced rates depend on the full set of two-subsystem coefficients $r_1^{(2)}, r_2^{(2)}$, which now contain the pair coherence $\rho_{nd}$. Increasing $\lambda$ enhances the coherent mixing between $|eg\rangle$ and $|ge\rangle$, strengthening the collective interaction channel and suppressing the net upward transition rate relative to the downward one. The result is a monotonic decrease of $T_{\text{cav}}/T_{\text{atom}}$ with increasing $\lambda$, signaling a genuine quantum-enhanced refrigeration mechanism unavailable in the one-subsystem configuration.

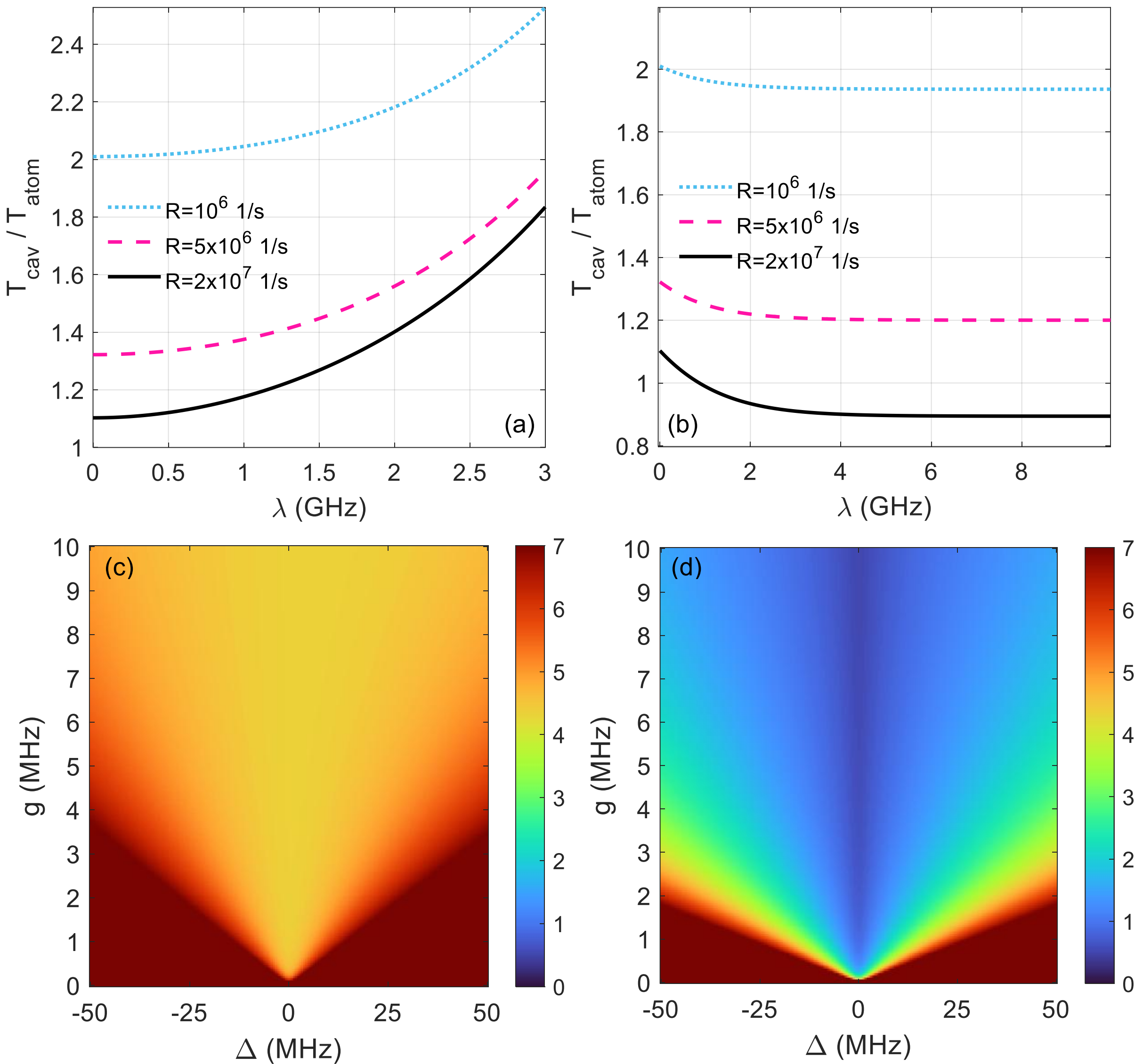

Figure 4: *(a,b) Influence of the intra-pair exchange coupling $\lambda$ on cavity cooling.* The ratio $T_{\mathrm{cav}}/T_{\mathrm{atom}}$ is plotted as a function of the exchange coupling $\lambda$ for the one-atom (a) and two-atom (b) interaction models. The exchange term $\lambda$ reshapes the correlated pair's internal populations $(\rho_e, \rho_g, \rho_d, \rho_{nd})$ and thereby modifies the stream-imposed detailed balance acting on the cavity. In the one-atom model (a), increasing $\lambda$ raises $T_{\mathrm{cav}}/T_{\mathrm{atom}}$, reflecting the reduced ability of the correlated pair to extract entropy when only a single subsystem interacts with the cavity. In contrast, in the two-atom model (b), stronger exchange enhances collective coherence and improves the net cooling power, leading to a monotonic decrease of $T_{\mathrm{cav}}/T_{\mathrm{atom}}$ with $\lambda$. *(c,d) Two-dimensional cooling landscape versus detuning and coupling strength.* Panels (c) and (d) show color maps of $T_{\mathrm{cav}}/T_{\mathrm{atom}}$ as a function of both the detuning $\Delta$ and the atom-cavity coupling strength $g$, for the one-atom and two-atom models, respectively. Cooling "valleys" appear near resonance and deepen with increasing $g$, while off-resonant or weak-coupling regions are dominated by phonon-induced heating. The two-atom model (d) exhibits a markedly stronger refrigeration effect and a broad region where $T_{\mathrm{cav}}/T_{\mathrm{atom}} < 1$, indicating true cavity cooling below the atom-stream temperature. These maps delineate the operational regime in which correlated atomic pairs can overcome phonon heating and impose a colder nonequilibrium steady state on the cavity.

Figure 4 (c) and (d) present two-dimensional maps of $T_{\mathrm{cav}}/T_{\mathrm{atom}}$ as functions of detuning $\Delta$ and coupling strength $g$. These maps synthesize, in a single representation, the competing influences of spectral overlap and interaction strength. Near resonance, spectral matching is optimal and increasing $g$ amplifies each weak collision, producing the "cooling valleys" characteristic of effective stream-dominated refrigeration. Far from resonance, the Lorentzian overlap $L(\Delta)$ suppresses stream-induced exchange, and phonon-induced heating drives the cavity toward $T_{\mathrm{bath}}$, generating broad heating plateaus at large $|\Delta|$.

The contrast between Figure 4 (c) and (d) reinforces the fundamental distinction between the one-atom and two-atom models: correlations play only a minor role when one subsystem interacts with the cavity, but become central when both atoms couple collectively. In the two-atom landscape, the enhanced coherence channels produce a wide region where $T_{\mathrm{cav}}/T_{\mathrm{atom}} < 1$, highlighting a robust refrigeration regime not accessible to the one-subsystem.

### 3.4 The effect of phonon damping

Figure 5 directly illustrates how cavity-phonon dissipation reshapes the steady-state temperature in both the one-atom and two-atom interaction models. When the cavity-phonon damping rate $\kappa \to 0$, the cavity becomes effectively isolated from its environmental phonon bath, and its temperature is governed almost entirely by the correlated atomic reservoir. In this limit the one-atom model saturates at the characteristic value $T_{\mathrm{cav}}/T_{\mathrm{atom}} \sim 4$, while the two-atom model reaches the substantially colder value $T_{\mathrm{cav}}/T_{\mathrm{atom}} \sim 0.5$. These limiting ratios reproduce the predictions of the idealized $\kappa = 0$ theory and serve as reference baselines for evaluating all realistic scenarios.[19]

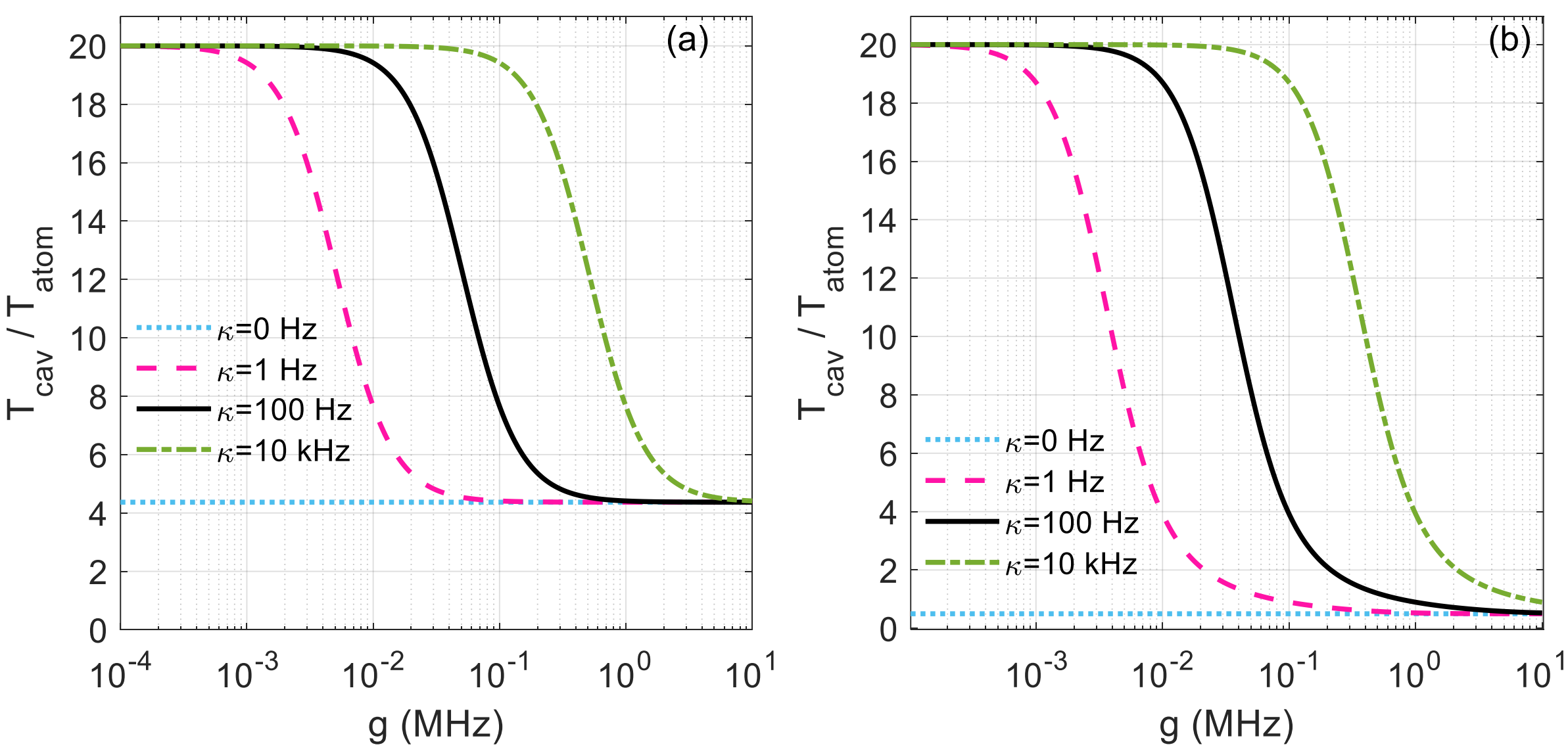

Figure 5: *Influence of cavity–phonon damping on cavity cooling.* The ratio $T_{\mathrm{cav}}/T_{\mathrm{atom}}$ is plotted for several cavity-phonon damping rates $\kappa$ in the one-atom (a) and two-atom (b) interaction models. These curves illustrate the crossover between the idealized limit $\kappa \to 0$, where the cavity is effectively isolated from its phonon environment, and the realistic phonon-tethered regime relevant for superconducting microwave cavities. In the weak-phonon-coupling limit, the cavity decouples from the environmental bath and the steady state is governed almost entirely by the correlated atomic reservoir. In this regime the one-atom model saturates near $T_{\mathrm{cav}}/T_{\mathrm{atom}} \sim 4$, whereas the collective two-atom model reaches significantly colder values $T_{\mathrm{cav}}/T_{\mathrm{atom}} \sim 0.5$, reflecting its enhanced refrigeration capability. As the damping rate $\kappa$ increases, phonon-induced excitations progressively suppress the stream's ability to extract entropy, and the cavity relaxes toward the phonon-bath temperature $T_{\mathrm{bath}}$. The sensitivity of these curves to $\kappa$ demonstrates how even modest phonon leakage reshapes the cooling window and establishes practical limits on stream-dominated refrigeration in realistic superconducting cavity platforms.

As soon as $\kappa$ becomes nonzero, even at values as small as a few hertz, phonon-induced excitations begin to compete with the entropy extraction performed by the atomic stream. This additional thermalization pathway progressively pulls the cavity toward the phonon-bath temperature $T_{\mathrm{bath}}$, weakening the stream-imposed detailed balance. The resulting crossover is clearly visible in both panels of Figure 5: for increasing $\kappa$, the cavity temperature rises from its stream-dominated value toward the phonon-dominated equilibrium. Because the one-atom model begins from a relatively hot baseline, its temperature rises more rapidly, whereas the two-atom model retains its refrigeration advantage over a broader window of phonon damping.

These trends delineate the practical regime in which engineered reservoir cooling can outperform environmental dissipation. Even modest phonon damping significantly narrows the operational window for strong cooling, emphasizing the importance of high-quality superconducting cavities for realizing stream-driven refrigeration. At the same time, the two-atom model demonstrates a substantially greater robustness to phonon leakage, reflecting the enhanced cooling power generated by pair coherence and collective interaction.

Thus, Figure 5 confirms both the consistency of our model with the $\kappa \to 0$ limit and the necessity of incorporating phonon coupling when assessing cooling performance in realistic experimental platforms.

A noteworthy distinction between the one-atom and two-atom coupling models becomes evident when examining the detuning dependence in the idealized limit of vanishing cavity-phonon contact ($\kappa \to 0$), as shown in Figure 6. In the one-atom configuration, the ratio $T_{\mathrm{cav}}/T_{\mathrm{atom}}$ displays a weak but discernible asymmetry as the atomic transition $\omega = \omega_1 + \Delta$ is swept over the range $\Delta \in [-50,50]$ MHz. This asymmetry arises from the unequal thermal weights $\rho_e \propto e^{-\beta\hbar\omega}$ and $\rho_g \propto e^{+\beta\hbar\omega}$, which feed directly into the detailed balance through the one-subsystem arrival coefficients $r_1$ and $r_2$. Because only a single spin interacts with the cavity, these coefficients retain the antisymmetric dependence on $\Delta$, leading to a slow monotonic drift of $T_{\mathrm{cav}}/T_{\mathrm{atom}}$ as the atomic frequency is tuned.

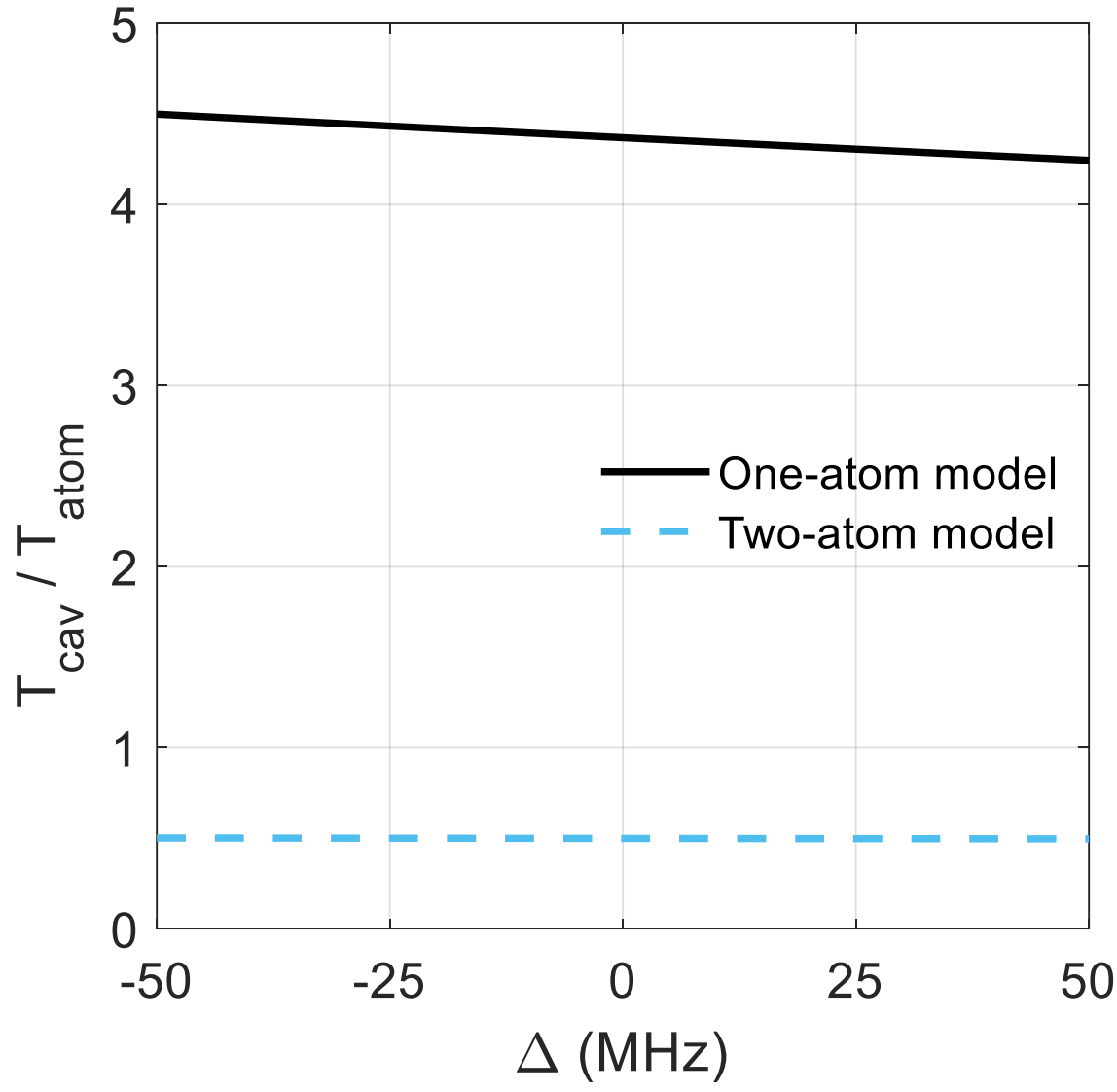

Figure 6: *Effective cavity temperature in the one- and two-atom interaction models in the absence of phonon coupling.* The ratio $T_{\mathrm{cav}}/T_{\mathrm{atom}}$ is shown as a function of the detuning $\Delta = \omega - \omega_1$ for $\kappa = 0$, isolating the intrinsic effect of the correlated atomic reservoir. In the one-atom model, the curve exhibits a weak asymmetry as Δ is swept over $\pm 50$ MHz: $T_{\mathrm{cav}}/T_{\mathrm{atom}}$ decreases approximately linearly with increasing detuning and continues to fall even for $\Delta > 0$. This slight asymmetry originates from the unequal thermal weights $\rho_e$ and $\rho_g$ that determine the effective upward and downward transition rates when only a single atom couples to the cavity. The two-atom model displays a much weaker dependence on Δ. Because both atoms interact collectively with the cavity, the relevant upward–downward rates incorporate symmetric pair-coherent contributions that suppress the detuning-induced imbalance present in the one-atom configuration. As a result, the two-atom curve is nearly symmetric about $\Delta = 0$, reflecting the enhanced robustness of collective cooling against detuning.

In contrast, when both atoms interact with the cavity, the correlated contributions from the single-excitation manifold, encoded in $\rho_d$ and, crucially, the coherence term $\rho_{nd}$, modify the effective upward and downward transition rates. These collective terms suppress the antisymmetric Δ-dependence present in the one-atom case, producing an almost symmetric and significantly weaker detuning dependence. The two-atom model thereby exhibits a nearly flat response across the same detuning range.

This contrast highlights the deeper role played by pair correlations: even in the absence of any phonon bath, collective coupling alters the effective reservoir seen by the cavity, smoothing out detuning asymmetries that are otherwise unavoidable in single-spin interaction models.

## 4. Experimental Mapping to a cQED Architecture

### 4.1 Motivation and Cryogenic Context

A central challenge in scaling quantum hardware arises from the steep drop in available cooling power as one descends below the kelvin scale in standard dilution refrigerators. While the 20 mK stage typically offers only on the order of 10-20 μW, the 100 mK stage reaches sub-milliwatt levels (≈ 0.5 mW in modern systems), illustrating a multi-order-of-magnitude disparity across the cryostat tiers; representative specifications from commercial systems document ≈ 16 μW at 20 mK and ≈ 0.5 mW at 100 mK for LD-class units.[1] By contrast, auxiliary 1 K stages and dedicated 1 K systems can supply far larger cooling capacities, often at or above the ≳ 100 mW scale, making the 1-4 K envelope an attractive "high-power zone'' for co-locating electronics, filtering, and other infrastructure.[2]

This imbalance forces nearly all quantum devices (superconducting qubits and resonators, semiconductor spin qubits, low-noise amplifiers, and sensitive readout circuitry) to reside at the lowest-temperature plate, where cooling resources are scarcest and wiring heat loads become the dominant architectural constraint. Reviews of superconducting-qubit hardware emphasize these system-level pressures and the importance of cryogenic staging and thermal budgeting.[8]

A promising direction for alleviating this bottleneck is to relocate as much classical and passive hardware as possible to higher-temperature stages (1-4 K) while employing engineered quantum reservoirs to pull specific degrees of freedom, selected cavity modes, qubits, or mesoscopic devices, to effective temperatures far below the ambient of the supporting stage. The broader idea is rooted in quantum thermodynamics: few-qubit thermal machines and autonomous quantum absorption refrigerators can impose a desired detailed balance on a target mode without net external work, functioning as compact quantum coolers that create local regions of low entropy even inside a comparatively warm environment.[14,15] In parallel, rapid progress in cryo-CMOS electronics at the 4 K tier is validating the complementary system design strategy of shifting classical control and readout overhead to warmer plates, thereby reserving the coldest stage primarily for components that genuinely require millikelvin operation.[31,32]

Within this context, the two-atom (two-qubit) refrigerator architecture explored here functions as an engineered reservoir that can operate at a warmer stage yet impose a colder effective temperature on a target cavity mode or device degree of freedom. By combining the abundant cooling power available around 1 K with an autonomous quantum-refrigeration mechanism, this approach aims to relieve thermal budgets at the base plate while still preserving the low-entropy conditions required locally for high-fidelity operation.[3,10]

Figure 7 summarizes the circuit-QED implementation of the two-qubit refrigerator within a cryogenic environment. A high-Q 3D cavity, hosting the refrigerated mode at effective temperature $T_{\mathrm{cav}}$, is embedded in a phonon bath at $T_{\mathrm{bath}}$ and coupled coherently to a pair of superconducting qubits that constitute the engineered atomic reservoir at $T_{\mathrm{atom}}$. The schematic highlights the coherent Jaynes–Cummings exchange between the qubits and the cavity, the dissipative reset channel that returns the qubit reservoir to a low-entropy state via a Purcell-filtered decay path, and the resulting closed heat-flow loop in which entropy is repeatedly pumped from the cavity into the engineered reservoir and then into the 1 K stage. This layout makes explicit how a locally cold cavity mode can be sustained inside a comparatively warm but high-power cryogenic tier, setting the stage for the quantitative temperature profiles presented in Figure 8.

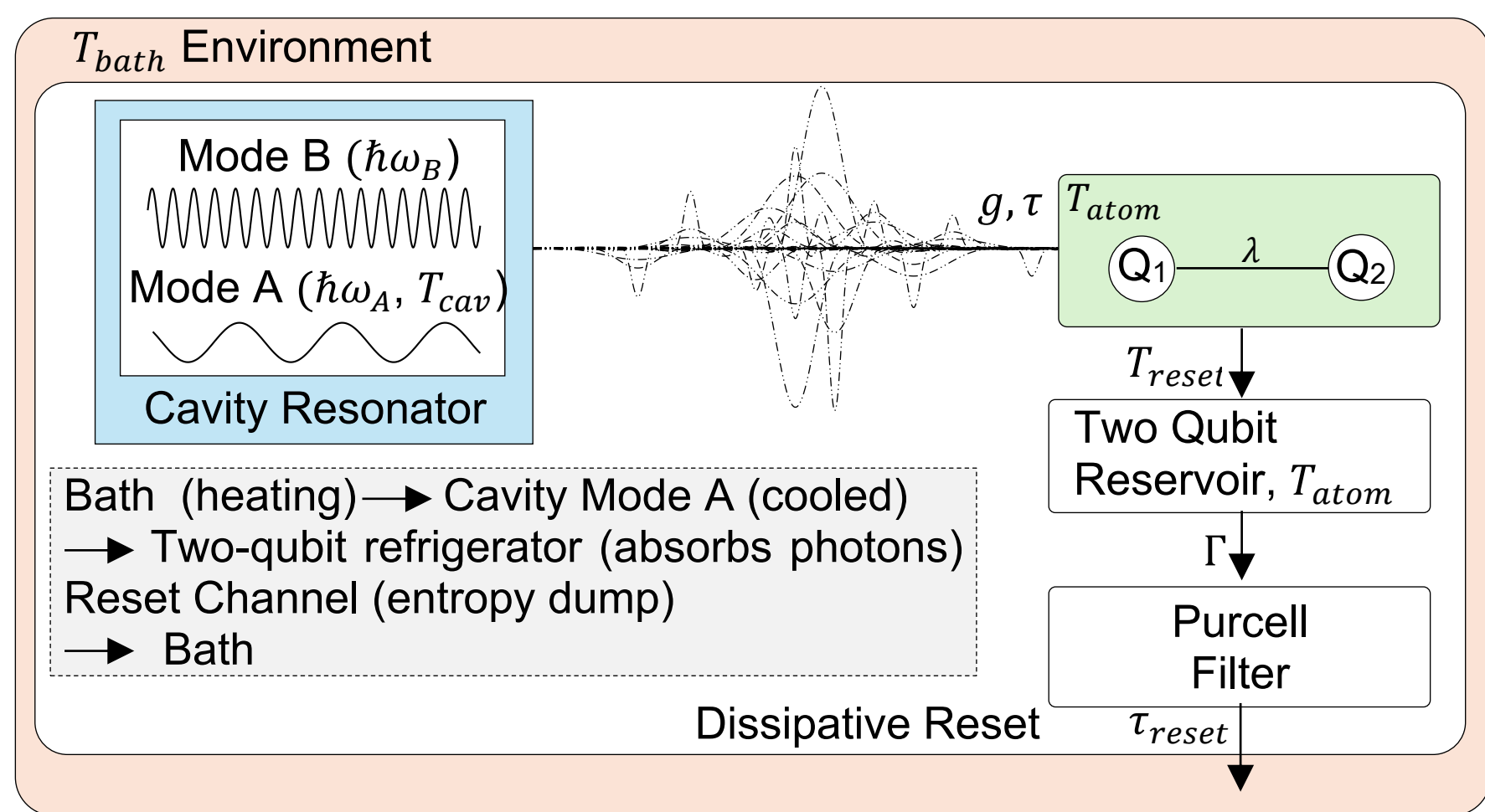

Figure 7: Circuit-QED implementation of the two-qubit refrigerator inside a cryogenic environment. A high-Q 3D cavity hosting Mode A at an effective temperature $T_{\mathrm{cav}}$ is embedded in a phonon bath at $T_{\mathrm{bath}}$ and coupled via Jaynes-Cummings interactions $g$ to a pair of superconducting qubits that constitute the engineered atomic reservoir at $T_{\mathrm{atom}}$. The arrow labeled $\Gamma$ represents a dissipative channel from the qubits into a coarse-grained two-qubit reservoir, which is reset through a Purcell-filtered decay path with reset time $\tau_{reset}$ into the $T_{\mathrm{bath}}$ stage. Together these processes realize a repeated-interaction refrigeration cycle that pumps entropy from the cavity into the engineered reservoir and ultimately into the bath, allowing the cavity mode A to equilibrate at $T_{cav}$ while residing on the same high-power cryogenic tier.

Figure 8 illustrates the practical significance of the engineered reservoir by showing the steady-state cavity temperature as a function of the cavity–bath damping rate $\kappa$ for several bath temperatures spanning the full cryogenic range relevant to superconducting hardware. In the one-subsystem case [Figure 8 (a)], the engineered reservoir drives the cavity to an effective temperature of $T_{\mathrm{cav}} \approx$ 219 mK in the limit of negligible bath contact, independently of $T_{\mathrm{bath}}$. Increasing $\kappa$ couples the mode more strongly to the environment, gradually pulling the cavity toward $T_{\mathrm{bath}}$: for a 100 mK bath the engineered refrigeration cools toward the bath temperature, while for higher bath temperatures (500 mK, 1 K, 4 K) the cavity instead warms toward the ambient. By contrast, the two-subsystem configuration [Figure 8 (b)] exhibits far stronger refrigeration: in the idealized limit $\kappa \to 0$, the cavity approaches $T_{\mathrm{cav}} \approx$ 25 mK, irrespective of the bath temperature. In the idealized collision model, pair quantum correlations enable the engineered reservoir to impose a substantially colder detailed balance on the cavity mode. In practice, the magnitude of this advantage depends on preserving the two-qubit coherence over the interaction window $\tau$.

In realistic devices, where $\kappa$ can be engineered to lie well below the kilohertz scale through careful cavity design and shielding, these results suggest that sub–100-mK effective cavity temperatures remain plausible, even when the surrounding cryostat is at 1 K or higher, provided parasitic dephasing and relaxation during each collision are sufficiently weak that the prepared two-qubit state is preserved over the interaction window. This establishes the feasibility of using a two-qubit refrigerator to create localized millikelvin regions inside high-power cryogenic stages, enabling cold microwave modes for quantum memories or as refrigeration interfaces for semiconductor qubits.

A practical reason to employ a two-atom (two-qubit) working medium is that direct “algorithmic” cooling of a bosonic cavity mode is fundamentally resource-intensive. A cavity has an infinite ladder of Fock states; cooling it algorithmically would require removing excitations at arbitrarily high photon numbers or equivalently engineering a highly selective dissipative channel that enforces detailed balance at a temperature below the ambient stage. Few-level systems, by contrast, admit fast, repeatable resets. In superconducting cQED, qubits can be reinitialized either autonomously by engineered dissipation or actively by measurement-based feedback, with mature experimental implementations. Fast unconditional all-microwave qubit reset and deterministic measurement-feedback reset have been demonstrated with sub-microsecond latencies, and driven-dissipative schemes autonomously stabilize target states, providing a compact route to entropy dumping.[23-25] These capabilities make qubits efficient *sacrificial refrigerants*, whereas the cavity serves as the cold stage that is repeatedly cooled through engineered contact with the qubit reservoir.

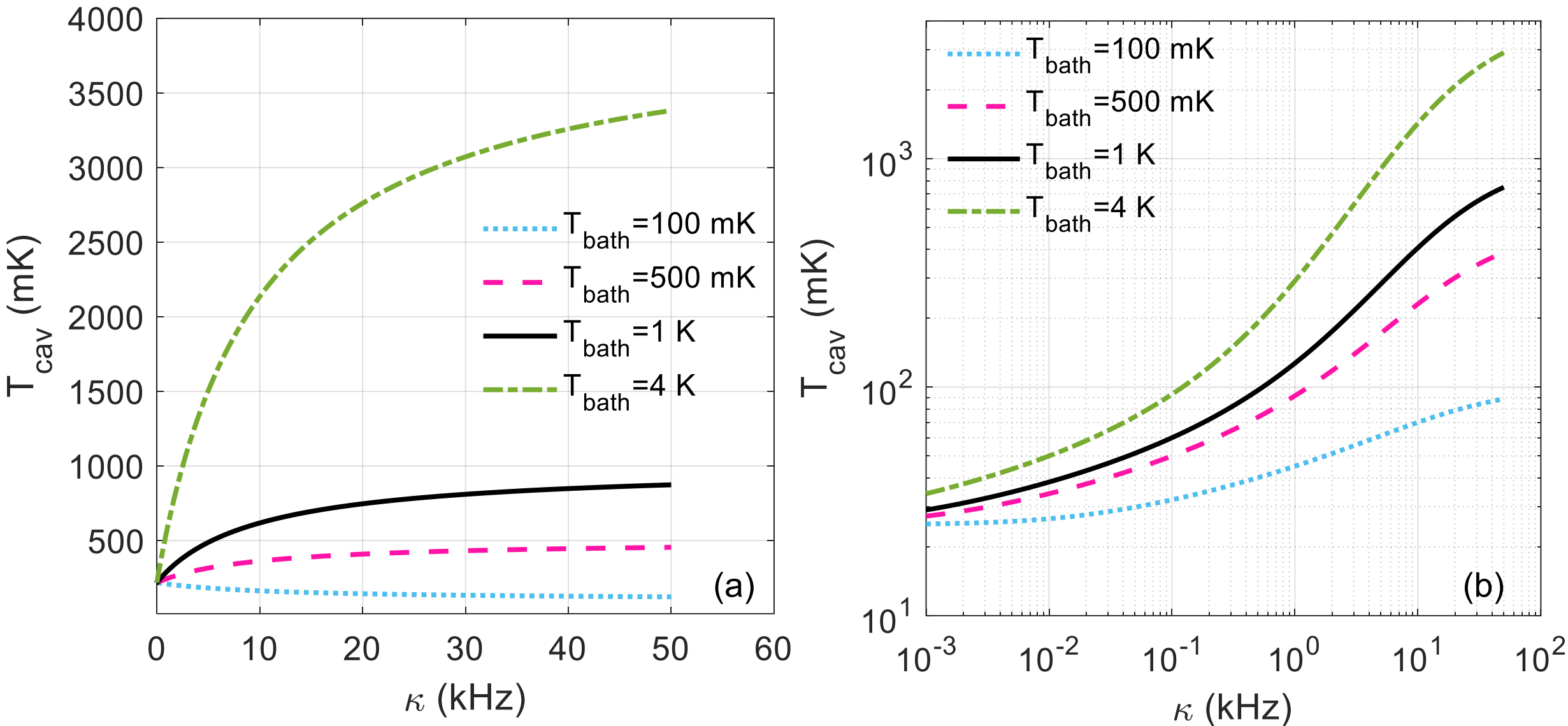

Figure 8: Steady-state cavity temperature $T_{\mathrm{cav}}$ versus cavity–bath damping rate $\kappa$ for (a) one-subsystem coupling (only one atom interacts with the cavity) and (b) two-subsystem coupling (both atoms interact). Curves are shown for four ambient bath temperatures $T_{\mathrm{bath}} =$ 100 mK, 500 mK, 1 K, 4 K. In (a), the one-atom configuration yields a minimum achievable cavity temperature of approximately $T_{\mathrm{cav}} \approx$ 219 mK at $\kappa = 0$, independent of $T_{\mathrm{bath}}$. As $\kappa$ increases, the cavity gradually equilibrates with the bath: at $T_{\mathrm{bath}} =$ 100 mK, the cavity cools toward 100 mK, while for higher $T_{\mathrm{bath}}$ it warms toward the bath temperature. In (b), the two-atom configuration enables significantly stronger refrigeration: for vanishing $\kappa$, the cavity reaches $T_{\mathrm{cav}} \approx$ 25 mK across all bath temperatures. Increasing $\kappa$ again drives the steady state toward $T_{\mathrm{bath}}$.

Using two qubits rather than one adds capabilities essential to our refrigerator. First, correlations in a two-qubit pair, embedded, for instance, in the singlet/triplet superpositions $|eg\rangle \pm |ge\rangle$, modify the upward and downward transition rates felt by the cavity. In the collision-model language, the coefficients $(r_1, r_2)$ or their two-subsystem extensions $(r_1^{(2)}, r_2^{(2)})$ are shifted by pair coherence, altering the effective detailed balance and enabling quantum-enhanced cooling, where the cavity's temperature $T_{\mathrm{cav}}$ drops below the effective temperature of the pair. This mechanism parallels collective emission and absorption in correlated emitters (Dicke super- and subradiance) and is closely related to micromaser-style cavity pumping by successive ancillae.[13,33,34]

Second, a two-qubit working medium broadens the design space for engineered reservoirs. By preparing and resetting appropriate two-qubit manifolds, one can tailor an effective temperature, spectral response, and energy-exchange directionality with the cavity, central objectives in quantum reservoir engineering and autonomous refrigeration.[16,35]

Finally, this mapping is directly compatible with modern hardware: the two-atom system can be realized with two transmons (planar or 3D), or with flux/fluxonium-type qubits, all within the standard circuit-QED toolset. Devices with stronger anharmonicity are especially attractive in the present context because they suppress thermal leakage out of the computational subspace at elevated-temperature stages. Strong and even ultrastrong light-matter couplings are routine, and tunable couplers provide wide, fast control of the qubit-qubit exchange needed to prepare and reset correlated states.[9,36-38] This establishes a clear experimental path for implementing the theoretical refrigerator analyzed in this work.

### 4.2 Realization with Two Superconducting Qubits and a 3D Cavity

The core elements of the proposed refrigerator, a high-Q microwave cavity and a pair of superconducting qubits, map naturally onto standard cQED hardware. A 3D superconducting cavity provides long photon lifetimes, clean electromagnetic mode structure, and compatibility with tunable and fixed-frequency qubits. In this architecture, the cavity plays the role of the cold working medium, while the two qubits act as a resettable quantum reservoir that repeatedly absorbs entropy from the cavity and dumps it into a cold load through active reset.

Each cooling cycle begins with preparing the two qubits in a prescribed mixed state with well-defined populations and pair coherence $(\rho_e, \rho_g, \rho_d, \rho_{nd})$. This can be achieved with a short sequence of single-qubit rotations and an entangling gate, optionally combined with engineered dissipation to stabilize the desired manifold. Such driven-dissipative state-preparation schemes are standard in quantum reservoir engineering, which uses the competition between coherent drive and controlled dissipation to autonomously stabilize target subspaces or entangled states.[23-25]

These preparation procedures realize the "initial ancilla state'' assumed in the collision-model section of our theory. The coarse-grained Lindbladian emerging from repeated ancilla-cavity collisions is rigorously justified in the open-quantum-systems literature,[11] providing the microscopic basis for describing each engineered two-qubit pair as a single "collision" imparting an $\mathcal{O}(\phi^2)$ kick to the cavity, where $\phi = g\tau$ is the Jaynes-Cummings interaction angle.

To implement the interaction time $\tau$ assumed in the theory, the qubits are tuned into (near) resonance with the cavity for the desired interval by applying nanosecond-scale flux pulses. Experimental demonstrations across both 3D and planar cQED platforms show that qubit transition frequencies can be shifted by hundreds of MHz with sub–100-ns latencies, enabling rapid "engage" and "disengage'' operations.[39,40] Tunable-coupler architectures and flux-tunable transmons routinely provide multi-hundred-MHz to multi-GHz frequency agility, allowing precise control of both the qubit-cavity detuning $\Delta$ and the inter-qubit exchange coupling $\lambda$.

During the engaged interval $\tau$, the two-qubit system and the cavity undergo a weak Jaynes-Cummings exchange. By choosing $\phi = g\tau \ll 1$, a single interaction reproduces the weak-collision limit required in the theoretical model, including the detuning-filtered energy exchange rate governed by the Lorentzian overlap $L(\Delta)$. Experimental control of detuning is routine in cQED, with MHz- to GHz-scale offsets readily programmed by flux modulation.

A practical limitation of the two-qubit reservoir is decoherence of the correlated pair during the engaged interval. In the idealized theory, the two-subsystem coefficients entering the cavity dynamics are $r_1^{(2)} = \rho_e + \rho_d + \rho_{nd}$ and $r_2^{(2)} = \rho_g + \rho_d + \rho_{nd}$, so the refrigeration advantage depends explicitly on the survival of the off-diagonal coherence $\rho_{nd}$. A simple phenomenological way to account for pair dephasing during the interaction window $\tau$ is therefore to replace $\rho_{nd}$ by an effective value $\rho_{nd}^{eff} = \rho_{nd}\, exp(-\tau/T_{2,pair})$, where $T_{2,pair}$ denotes the coherence time of the prepared two-qubit manifold during the collision. The effective stream coefficients then become $r_{1,eff}^{(2)} = \rho_e + \rho_d + \rho_{nd}^{eff}$ and $r_{2,eff}^{(2)} = \rho_g + \rho_d + \rho_{nd}^{eff}$. Because $\rho_{nd} < 0$ for the exchange-coupled thermal pair considered here, decay of the off-diagonal element weakens the quantum enhancement and shifts the cavity steady state toward a warmer, more nearly incoherent limit. In the extreme case $\tau \gtrsim T_{2,pair}$, the two-qubit advantage is largely erased. By contrast, when $\tau \ll T_{2,pair}$, the coherence reduction during a single collision is small and the idealized analytic results remain a good approximation.

If the qubits are coupled to the same thermal environment as the cavity, additional relaxation and dephasing during the collision will similarly renormalize both the pair populations and coherence; the protocol therefore remains most favorable in the regime $\Gamma_1\tau \ll 1$ and $\Gamma_2\tau \ll 1$, where the ancilla state is only weakly perturbed during each interaction window.

After interaction with the cavity, both qubits must be returned rapidly to a low-entropy state before the next cycle. Reset is the thermodynamic "work step'' that evacuates the entropy absorbed from the cavity. Circuit-QED experiments now support several mature reset modalities, all compatible with MHz-rate refrigeration cycles:

1. Dissipative (Purcell-enhanced) reset:
   The qubit is temporarily coupled to a low-Q mode or Purcell-filtered decay channel, enabling unconditional relaxation with time constants of order $10^2$ ns and residual excitations at or below $10^{-3} - 10^{-2}$.[26-29]
2. Measurement-and-feedback reset:
   High-fidelity projective measurement followed by a conditional $\pi$-pulse yields near-ground-state preparation on microsecond timescales and with high robustness[24,41]
3. Autonomous (driven-dissipative) reset:
   Parametric or multi-tone microwave drives engineer an attractor in the qubit's state space, enabling unconditional ground-state reset in sub–500-ns without classical feedback.[42,43]
4. Stimulated-emission and parametric-sideband reset:
   Recently demonstrated architectures use broadband or parametrically driven couplings to funnel qubit excitations into a cold bath with sub–100-ns latencies while suppressing unwanted Purcell decay during idle periods.[38,44]

These reset modalities have been extensively optimized in the cQED community and reliably maintain effective qubit temperatures in the 30-80 mK range at 5-7 GHz transition frequencies, consistent with residual populations of ~1-3%. Because the 1-K stage provides abundant cooling power, the classical microwave resources needed to drive, measure, and reset the qubits are easily supported at that temperature.

Once preparation, interaction, and reset are available, the full repeated-interaction refrigerator is straightforward to realize. Each "cooling cycle'' consists of:

1. Prepare correlated two-qubit state.
2. Bring both qubits into resonance with the cavity for time $\tau$.
3. Allow an $\mathcal{O}(\phi^2)$ energy exchange event with adjustable detuning Δ.
4. Disengage and reset both qubits.
5. Repeat at a programmable rate $R$.

Randomizing the start time of each cycle (or using an aperiodic digital trigger) reproduces the Poissonian arrival statistics underlying the collision model and the master equation used in sections 1 and 2. This mapping has deep roots in the micromaser literature, where successive atoms passing through a cavity generate an effective Lindblad generator with upward and downward rates proportional to $Rr_1\phi^2$ and $Rr_2\phi^2$.[13] Under realistic parameters (MHz-class arrival rates, small angles $\phi \lesssim 0.1$, detunings in the 10-100-MHz range, and cavity damping 10-100 kHz) the resulting effective damping $\Gamma_\downarrow$ and source term $J_\uparrow$ yield cavity-cooling time constants of order 10-100 µs, compatible with standard cQED coherence windows.

Thus, the repeated-interaction refrigerator studied in our theory can be implemented almost directly using off-the-shelf cQED components. All required ingredients (fast state preparation, rapid tunability, high-fidelity reset, and multimode cavity engineering) are already experimentally demonstrated, establishing the feasibility of recreating our idealized two-atom model within present-day superconducting-qubit hardware.

### 4.3 Reset Mechanisms and Achievable Temperatures

Reset plays a central role in enabling the two-qubit refrigerator, as it constitutes the entropy-dumping step required to maintain the working medium at low effective temperature. Without a fast and high-fidelity return to a low-entropy state, the qubits would accumulate excitations extracted from the cavity and the refrigeration cycle would saturate. Within circuit QED, qubit reset has matured into a well-developed technological capability, with several experimentally established modalities that operate at rates compatible with the MHz-class repeated-interaction cycles assumed in our theoretical model. These reset channels exploit either engineered dissipation or active measurement and control to funnel population out of the excited state and thereby maintain the qubit subsystem at a low effective temperature.

Dissipative reset is often realized through Purcell-enhanced relaxation, where the qubit is temporarily coupled to a low-Q auxiliary mode or a dedicated Purcell filter so that the excited state decays rapidly into a cold load. Experiments have demonstrated unconditional ground-state reset with characteristic time constants on the order of $10^2$ ns and with residual excited-state populations as low as $10^{-3} - 10^{-2}$; more recent architectures employ tunable or multi-mode Purcell filters to combine rapid reset with protection against unwanted decay during coherent operations.[26,29,37,38] A complementary approach uses measurement followed by classical feedback, wherein a high-fidelity projective measurement is performed and an appropriate conditional $\pi$-pulse is applied to return the qubit to its ground state. Deterministic, measurement-based reset routinely achieves near-ground-state preparation within microsecond time intervals and has enabled repeated initialization well beyond the limits set by intrinsic $T_1$ relaxation.[24,41]

Autonomous, measurement-free reset provides yet another path to rapid entropy removal. In such schemes, the competition between coherent microwave drive and engineered dissipation creates an attractor in the qubit state space that deterministically funnels population toward the ground state, eliminating the need for feedback loops. Driven-dissipative protocols have prepared fixed-frequency transmons to their ground state within sub–500-ns windows and have been extended to the unconditional stabilization of entangled two-qubit states, making the approach particularly appealing for thermal machines based on correlated qubit pairs.[23,42,43] Related ideas activate stimulated emission or parametric sideband transitions to accelerate population transfer into a lossy bath. Architectures employing broadband waveguides or parametrically modulated couplers have reported unconditional reset latencies below 100 ns, together with the ability to suppress Purcell decay during idle times through tunable coupling elements.[27,38]

These reset modalities are powerful enough to maintain the two-qubit subsystem at an effective temperature in the $30$-$80$ mK range even while operating at a 1 K cryostat stage. A residual excited-state probability of only a few percent at qubit transition frequencies in the 5-7 GHz range corresponds directly to effective temperatures of this order, consistent with the targets needed for the refrigerator. The classical microwave resources required to implement these resets (high-power drives, broadband control lines, and fast measurement electronics) can be supported readily at the 1 K tier, which typically offers cooling powers in the hundreds of milliwatts. Commercial 1 K platforms provide between 200-700 mW at temperatures around 1-1.2 K, and dedicated 1 K-loop stages integrated into dilution refrigerators offer similarly high capacities.[1,5] These conditions make the 1 K environment particularly attractive for hosting the qubit-reservoir infrastructure, with the cavity mode acting as the object to be cooled inside this warm but powerful stage.

A related practical issue is leakage out of the ideal two-level manifold when the superconducting ancilla is hosted at an elevated cryostat stage. If the qubits were allowed to equilibrate passively with a 1–4 K environment, higher excited states would indeed acquire substantial population, and the working medium would have to be treated as a multilevel system rather than as an effective qubit pair. The present proposal therefore relies essentially on active reset and state preparation to define a low-entropy effective two-level manifold prior to each collision. When the reset maintains an effective ancilla temperature in the 30–80 mK range at $\omega_{01}/2\pi \sim 5$–7 GHz, the first excited-state population remains only a few percent, while the population of the next level is further suppressed by the additional Boltzmann factor associated with the $1 \rightarrow 2$ transition. For a transmon, this makes leakage to $|f\rangle$ small provided the anharmonicity remains large compared with the relevant linewidths and detunings, so that the cavity interacts predominantly with the $0 \leftrightarrow 1$ transition. In this regime, higher levels mainly reduce the effective reset fidelity and slightly renormalize the prepared pair state, thereby weakening the refrigeration enhancement rather than qualitatively changing its mechanism. If higher-level occupations become non-negligible, however, the ancilla must be described by a qutrit- or multilevel-reservoir model with additional upward and downward transition channels; such channels would generally introduce extra heating pathways and reduce the net cooling power. From an implementation perspective, this consideration favors strongly anharmonic devices such as fluxonium, or transmon protocols with explicit leakage-reduction and reset calibration, when operating at elevated-temperature stages.

The resulting steady-state temperatures predicted by our collision-model dynamics reflect this balance. Under realistic conditions (MHz-scale repetition rates, small interaction angles $\phi = g\tau$, weak cavity-bath coupling $\kappa$ in the 10-100 Hz to 1 kHz range, and detuning-controlled exchange governed by the spectral filter $L(\Delta)$) the cavity cools toward a steady state set predominantly by the engineered qubit reservoir rather than by the ambient 1 K bath. The resulting cavity temperature $T_{\mathrm{cav}}$ lies in the 50-120 mK range across broad regions of parameter space, and the effective cooling time constants fall in the $10\text{-}100\,\mu s$ band, consistent with the $O(R\phi^2)$scaling of repeated-interaction refrigerators and with the coherence times accessible in high-Q 3D cavities.

The multimode nature of 3D cavities further expands the utility of these reset mechanisms. When a second cavity mode (Mode B) is used as a memory or as an interface to a semiconductor quantum dot, it can be cooled opportunistically, only when it stores no quantum information, by temporarily activating a parametric beam-splitter interaction $g_{AB}(t)$ that swaps excitations with the refrigerated Mode A. Recent demonstrations of high-isolation, high-fidelity parametric mode converters, including SNAIL-based and parity-protected devices, confirm that such inter-cavity beam-splitters can be activated with large on-off ratios and low added loss.[45,46] In this "cool-before-load'' protocol, entropy flows from Mode B into Mode A and then into the engineered two-qubit reservoir through the reset channel, leaving Mode B cold and ready to store quantum information or couple to an external device.

Direct algorithmic cooling of a harmonic oscillator, by contrast, remains fundamentally difficult. Because of the cavity's infinite-dimensional Hilbert space, removing population from arbitrarily high Fock states would require non-Gaussian, number-resolved operations or photon-number-specific measurements that are experimentally demanding even in strongly dispersive cQED architectures. The literature on heat-bath algorithmic cooling (HBAC) targets finite-dimensional registers and consistently relies on rapidly resettable ancilla qubits as the fundamental resource for entropy compression.[11,40] General constraints on Gaussian thermal operations in bosonic systems reinforce this limitation, as achieving very low effective temperatures requires either non-Gaussian resources or auxiliary systems that can be efficiently reset, returning us to the two-qubit reservoir as the minimal realistic module for cooling a cavity mode.[30,40,47]

Taken together, these developments establish that the two superconducting qubits used in our refrigerator can reliably be reset to effective temperatures in the tens of millikelvin range at rates exceeding those required for the repeated-interaction dynamics. They therefore provide exactly the type of low-entropy, rapidly recyclable ancilla needed for enforcing a cold detailed balance on a cavity mode situated within a high-power 1 K environment.

## 5. Outlook

The experimental analogue of the theoretical "two-atom refrigerator" is a pair of actively reset superconducting qubits that repeatedly interact with a high-Q 3D microwave cavity mode. In this picture the qubits function as a sacrificial, low-entropy reservoir maintained far below the ambient cryostat temperature by fast dissipative or measurement-based resets; their repeated, weak, and detuned-tunable collisions impose an effective detailed balance on the cavity, cooling it to tens of millikelvin even when the cryostat itself operates near 1 K. This is precisely the regime in which 1 K platforms provide abundant cooling power, hundreds of milliwatts to the watt class, so the entropy removal required for frequent qubit resets can be supplied without stressing the base-plate budget, while the cavity experiences a tailored quantum reservoir rather than the warm phonon bath. The repeated-interaction framework gives a microscopically controlled route from discrete ancilla shots to a Lindbladian for the cavity, with the micromaser literature and modern collision-model theory offering both the conceptual and quantitative underpinnings for this mapping.[13]

Near-term milestones are within reach. Fast, high-fidelity qubit resets at the 100-ns scale, via Purcell-enhanced dissipation, parametric channels, or autonomous reservoir engineering, are now routine, and autonomous entanglement stabilization of two-qubit manifolds has been demonstrated in cavity- and waveguide-based settings. In our architecture these elements provide the "reset and reprepare" primitive that clamps the ancilla pair to a prescribed mixed state with controllable correlations, which in turn fixes the upward/downward coefficients in the collision map and therefore the cavity's steady-state temperature. Experimentally, one can validate the cooling by cavity thermometry (e.g., sideband spectroscopy on a weak probe qubit or mode) and by measuring the ancilla energy flow during cycles; both readouts are compatible with MHz repetition and small-angle exchanges.[48,49,50]

Once established, a precooled cavity enables device-level payoffs. Gate-defined semiconductor double quantum dots (DQDs)[51] have reached strong coupling with microwave cavities in III-V and Si/SiGe platforms; in such devices a colder cavity directly suppresses thermal photon occupation at the qubit transition and reduces phonon-assisted leakage to unwanted orbitals, improving spin-qubit initialization and charge-noise resilience at elevated cryostat temperatures. A second, complementary pathway is inter-cavity refrigeration: designate one cavity mode as the refrigeration interface and use a programmable beam-splitter to swap entropy from a memory or DQD-coupling mode when it is empty ("cool-before-load"), a control primitive already standard in bosonic-code experiments. Together these uses make the refrigerated cavity a versatile low-entropy resource embedded in a warm 1 K environment. [52-54]

From a systems perspective, the architecture addresses a central scalability bottleneck: millikelvin stages offer only tens of microwatts at the mixing chamber, whereas 1 K stages can sustain two to three orders of magnitude more power for control electronics, pumps, and fast flux drives. Relocating control to 1 K while generating local sub–100-mK pockets via qubit-engineered reservoirs provides a realistic path toward larger channel counts without saturating the base plate. Theoretical control is aided by the detuning-aware collision model: the same small-angle interaction can be spectrally gated by $L(\Delta)$, allowing robustness to drift and deliberate trade-offs between interaction time and overlap bandwidth, while retaining a closed-form birth-death description for the cavity photon number. This separation (thermodynamics in the ancilla state, geometry in the two-atom enhancement, and spectroscopy in the detuning filter) offers clear handles for calibration and stability in the lab.[2,7]

Open technical challenges frame the near-term research agenda. Heat loads from rapid qubit resets and control pulses must be budgeted to avoid excess quasiparticles and stray photons; high-isolation, low-loss switching between "engage" and "idle" configurations is needed to maintain the small-angle regime; and Poissonization of cycle timing should be verified so the coarse-grained Lindblad limit applies. Multi-mode crowding in 3D enclosures can be turned from a nuisance into a feature by using engineered filters and parametric couplers to direct entropy flow, but this requires careful microwave design. On the metrology side, QND tools from circuit QED, micromaser-inspired counting statistics and modern bosonic-code diagnostics, can be repurposed to quantify the effective temperature and detailed balance imposed by the ancilla reservoir. Finally, integrating this refrigerator with error-corrected bosonic encodings (cat, binomial, GKP) suggests a route to hardware that both cools and stabilizes logical states at the 1 K tier, leveraging the same reservoir-engineering toolbox demonstrated in recent autonomous QEC experiments. [13,55]

In sum, by combining the high-power 1 K environment with a two-qubit, resettable working medium, the proposed refrigerator creates localized low-entropy regions, $T_{\mathrm{cav}}$ in the 50-120 mK range, without relying on the dilution refrigerator's base plate. The collision-model derivation, the experimental maturity of fast reset and autonomous stabilization, and the demonstrated strong coupling to semiconductor nanostructures collectively support a credible path to large-scale, higher-temperature quantum hardware in which refrigeration is delivered where and when it is needed, rather than being globally imposed by the coldest stage.

## Conclusion

We have developed a comprehensive theoretical framework for cooling a phonon-tethered microwave cavity using a stream of correlated two-level systems, including realistic phonon environments, detuning filters, finite cavity decay, and both single- and double-atom coupling configurations. By deriving explicit Lindblad-rate equations and analytically tractable steady-state solutions for the cavity photon number, we have shown how the interplay of weak collisions, intra-pair correlations, detuning, and phonon loss determines the nonequilibrium temperature imposed on the cavity. Our results demonstrate that while a single interacting atom in each correlated pair can partially suppress phonon-induced heating, genuine refrigeration, where the cavity is driven below the temperature of the atomic reservoir, requires the collective interaction of both atoms in the pair. This two-atom configuration harnesses exchange-enhanced correlations and coherence to reshape the effective detailed balance seen by the cavity, producing cooling minima as deep as $T_{\mathrm{cav}}/T_{\mathrm{atom}} \approx 0.5$ under realistic coupling strengths and spectral conditions.

Systematic evaluation of the analytic steady-state expressions across a wide parameter space reveals the operational boundaries of this quantum refrigerator, including its sensitivity to detuning, the crucial role of the cavity–bath leakage rate, and the enhancement provided by strong atom–cavity coupling. The resulting cooling landscapes identify broad regions in which the engineered atomic reservoir can overcome environmental heating and impose a colder nonequilibrium steady state. By comparing the one-atom and two-atom models, we have clarified the qualitative advantages of collective coupling, including stronger refrigeration, wider detuning tolerance, and reduced sensitivity to asymmetries in the thermal weights of the incoming states.

Motivated by these theoretical insights, we have mapped the two-atom cooling mechanism onto a practical cQED architecture. In this platform, two superconducting qubits serve as the working medium, a high-Q 3D cavity provides the target bosonic mode, and fast dissipative or measurement-based resets maintain the qubit pair at an effective temperature far below that of a 1 K cryogenic stage. This experimental mapping connects our theoretical model directly to state-of-the-art techniques in cQED, including Purcell-enhanced reset, parametric sideband interactions, tunable couplers, and intercavity beam-splitters. The combined system offers a realistic pathway for creating local sub–100-mK cold spots inside a warm enclosure, potentially enabling scalable quantum information architectures in which most control electronics reside at high-power cryogenic stages while only selected modes or devices are actively refrigerated via engineered quantum reservoirs.

Overall, this work bridges quantum thermodynamics, collision-model open-system theory, and modern cQED hardware to propose a feasible route toward autonomous cavity cooling far below ambient cryogenic temperatures. The enhanced performance of the two-atom reservoir underscores the central role of quantum correlations in nonequilibrium quantum machines and opens opportunities for integrating similar cooling modules with semiconductor quantum dots, spin qubits, bosonic memories, and other temperature-sensitive quantum devices. Future work may extend this framework to driven dissipative steady-state engineering, multi-mode refrigeration, and autonomous thermal management in large-scale quantum processors.

## Appendix A. Detuning kernel and limiting cases

### A.1 Origin of the overlap kernel

In this appendix we briefly show how the double-time integrals entering the detuning filter arise from the second-order collision map. For a single atom–cavity collision of duration $\tau$, the reduced cavity state is

$$\Phi(\rho) = \mathrm{Tr}_{\mathrm{atom}}\left[U_I(\tau)\,(\rho \otimes \rho_{\mathrm{atom}})\,U_I^{\dagger}(\tau)\right], \qquad \text{(A1)}$$

where $U_I(\tau)$ is the interaction-picture propagator generated by the Jaynes–Cummings interaction. Expanding $U_I(\tau)$ to second order in the small parameter $\phi = g\tau$, one obtains

$$U_I(\tau) \simeq I - \frac{i}{\hbar}\int_0^{\tau} dt\; H_{\mathrm{int}}(t) - \frac{1}{\hbar^2}\int_0^{\tau} dt \int_0^{t} ds\; H_{\mathrm{int}}(t) H_{\mathrm{int}}(s). \qquad \text{(A2)}$$

For the detuned one-atom interaction,

$$H_{\mathrm{int}}(t) = \hbar g(a\,\sigma_A^{+} e^{-i\Delta t} + a^{\dagger}\sigma_A^{-} e^{+i\Delta t}), \Delta \equiv \omega - \omega_1. \qquad \text{(A3)}$$

The first-order contribution vanishes after tracing over the incoming atom because the reduced atomic state is diagonal in the local energy basis, so that

$$\mathrm{Tr}_{\mathrm{atom}}(\sigma_A^{\pm}\rho_{\mathrm{atom}}) = 0. \qquad \text{(A4)}$$

Therefore, the leading nonvanishing correction to the cavity map is second order.

Substituting Eq. (A3) into the second-order Dyson terms produces products of the form

$$H_{\mathrm{int}}(t)\;(\rho \otimes \rho_{\mathrm{atom}})\;H_{\mathrm{int}}(s), \qquad \text{(A5)}$$

together with the corresponding anti-commutator terms. After tracing over the atomic degrees of freedom, only terms proportional to the local correlators

$$\mathrm{Tr}_{\mathrm{atom}}(\sigma_A^+\sigma_A^-\,\rho_{\mathrm{atom}}) = r_1, \mathrm{Tr}_{\mathrm{atom}}(\sigma_A^-\sigma_A^+\,\rho_{\mathrm{atom}}) = r_2 \qquad \text{(A6)}$$

survive, so the second-order correction has the Lindblad structure discussed in the main text. The oscillatory phases in Eq. (A3) combine to give factors $e^{\pm i\Delta(t-s)}$, which encode the atom–cavity detuning.

To incorporate the finite cavity linewidth induced by the phonon bath, we use the standard Markov cavity correlation

$$\langle a(t)a^\dagger(s)\rangle_{\mathrm{env}} \propto e^{-(\kappa/2)|t-s|}, \qquad \text{(A7)}$$

so that the relevant second-order kernel is weighted by both the detuning phase factor and the cavity correlation envelope. This yields double integrals of the form

$$\int_0^\tau dt \int_0^\tau ds \;\; e^{-(\kappa/2)|t-s|}e^{\pm i\Delta(t-s)}. \qquad \text{(A8)}$$

These are the integrals quoted in the main text. They quantify the spectral overlap between the finite interaction window and the damped cavity line.

Defining

$$I(\Delta) \equiv \int_0^\tau dt \int_0^\tau ds \;\; e^{-(\kappa/2)|t-s|}e^{i\Delta(t-s)}, \qquad \text{(A9)}$$

and changing variables to $u = t - s$, one obtains

$$I(\Delta) = 2\,\mathrm{Re}\int_0^\tau (\tau - u)\;\, e^{-(\kappa/2-i\Delta)u}\, du, \qquad \text{(A10)}$$

which is the kernel used in Sec. 1.2. Its limiting forms are discussed in Appendix A.2.

**A.2 Limiting forms of the detuning kernel**

The closed-form kernel in Eq. (14),

$$I(\Delta) = 2\,\mathrm{Re}\left\{\frac{\tau}{\alpha}(1-e^{-\alpha\tau}) - \frac{1}{\alpha^2}[1-e^{-\alpha\tau}(1+\alpha\tau)]\right\}, \alpha \equiv \frac{\kappa}{2} - i\Delta,$$

interpolates between two transparent limiting regimes.

In the first regime, $\kappa\tau \gg 1$, the interaction time is long compared to the cavity correlation time $2/\kappa$. The overlap is then cavity-limited, and the exponentially small terms in Eq. (14) can be neglected. One obtains

$$I(\Delta) \approx 2\left[\frac{\tau(\kappa/2)}{(\kappa/2)^2+\Delta^2} - \frac{(\kappa/2)^2-\Delta^2}{[(\kappa/2)^2+\Delta^2]^2}\right]. \qquad \text{(A11)}$$

The second term is independent of $\tau$, whereas the first term grows linearly with $\tau$. Accordingly, in the $\kappa\tau \gg 1$limit the leading behavior reduces to

$$I(\Delta) \approx \frac{\kappa\tau}{(\kappa/2)^2+\Delta^2}, \qquad \text{(A12)}$$

which is a Lorentzian function of $\Delta$with half width at half maximum $\kappa/2$. Thus, for long collisions the detuning dependence is governed primarily by the cavity linewidth.

In the opposite regime, $\kappa\tau \ll 1$, the cavity field is effectively constant during the short interaction window, and the kernel becomes time-window limited. In this case,

$$I(\Delta) \approx 2\int_0^\tau (\tau-u)\cos(\Delta u)\, du = \frac{\sin^2(\Delta\tau/2)}{(\Delta/2)^2} = \tau^2\,\mathrm{sinc}^2\left(\frac{\Delta\tau}{2}\right) \qquad \text{(A13)}$$

The resulting profile has a main lobe of characteristic width of order $1/\tau$, as expected from the Fourier transform of a finite interaction window.

These two asymptotic forms motivate the effective overlap filter introduced in the main text: at long interaction times the overlap is controlled by the cavity linewidth, whereas at very short interaction times it is controlled by the Fourier width of the collision window

**A.3 Consistency checks and parameter trends for the one-atom coupling configuration**

The detuning-dependent steady-state occupation derived in the main text,

$$n^*(\Delta) = \frac{\kappa\bar{n}_1 + Rr_1\phi^2 L(\Delta)}{\kappa + R(r_2 - r_1)\phi^2 L(\Delta)}, \qquad (24)$$

together with the corresponding effective cavity temperature,

$$T_{\text{cav}}(\Delta) = \frac{\hbar\omega_1}{k_B \ln\left(1 + 1/n^*(\Delta)\right)}, \qquad (26)$$

admits several useful limiting cases that help clarify the physical content of the one-atom coupling configuration.

First, in the resonant limit $\Delta = 0$, the overlap factor satisfies $L(0) = 1$. Equations (23)–(26) then reduce to

$$n(t) = n^* + [n(0) - n^*]e^{-\Gamma_\downarrow t}, \qquad n^* = \frac{\kappa\bar{n}_1 + Rr_1\phi^2}{\kappa + R(r_2 - r_1)\phi^2}, \qquad (A14)$$

with effective cavity temperature

$$T_{\text{cav}} = \frac{\hbar\omega_1}{k_B \ln\left(1 + 1/n^*\right)}. \qquad (A15)$$

Thus, the detuned theory reduces smoothly to the resonant result when the atom and cavity are spectrally aligned.

Second, when the stream is effectively switched off, $R\phi^2 \to 0$, the engineered reservoir no longer contributes to the photon-number dynamics. In that limit,

$$n^*(\Delta) \longrightarrow \bar{n}_1,$$

so that the cavity thermalizes to the ambient bath temperature $T_{\text{bath}}$, as expected.

Third, in the opposite limit of vanishing cavity-bath coupling, $\kappa \to 0$, the steady-state cavity occupation becomes

$$n^*(\Delta) \underset{\kappa\to 0}{\to} \frac{r_1}{r_2 - r_1}. \qquad (A16)$$

This limit is independent of detuning because the same overlap factor $L(\Delta)$ appears in both the numerator and denominator of Eq. (24) and therefore cancels identically. On resonance, this reproduces the one-atom stream relation

$$\frac{n^*}{n^* + 1} = \frac{r_1}{r_2},$$

which may be rewritten as

$$e^{-\hbar\omega_1/(k_B T_{\text{cav}})} = \frac{r_1}{r_2} = \frac{e^{-\beta\hbar\omega} + \cosh\left(\beta\hbar\lambda\right)}{e^{+\beta\hbar\omega} + \cosh\left(\beta\hbar\lambda\right)}, \beta = \frac{1}{k_B T_{\text{atom}}}. \qquad (A17)$$

Equation (A17) makes explicit how the cavity temperature is set by the effective detailed balance of the atom stream when the phonon bath is absent.

Another useful regime is that of large detuning or very short interaction time. For $|\Delta| \gg \Gamma_{\text{over}}$, the overlap factor satisfies $L(\Delta) \ll 1$, so the influence of the stream is strongly suppressed and $n^*(\Delta)$approaches $\bar{n}_1$. In this sense, large detuning effectively disconnects the engineered reservoir from the cavity. For very short collisions, the Fourier width $1/\tau$ dominates the effective overlap width $\Gamma_{\text{over}} = \kappa + 1/\tau$, and the detuning filter reduces approximately to

$$L(\Delta) \approx \frac{1}{1 + (2\Delta\tau)^2}$$

This reveals a useful trade-off: the overall stream strength grows as $\phi^2 = g^2\tau^2$, but detuning is penalized less strongly for shorter interaction times because the interaction window broadens spectrally. Consequently, for fixed $g$, $\Delta$, and $\kappa$, there is generally an optimal interaction time $\tau$that maximizes the stream's net leverage on the cavity.

Finally, it is useful to emphasize the role of the intra-pair exchange coupling $\lambda$. The detuning filter $L(\Delta)$ modifies only the overall strength of the stream-induced pump and damping; it does not affect their bias, which remains governed by $r_2 - r_1$. Through Eqs. (2)–(7), this bias depends on $(\omega, T_{\text{atom}}, \lambda)$. Increasing $|\lambda|$ increases the partition function $Z$, thereby reducing the asymmetry $r_2 - r_1$ and weakening the net cooling or heating leverage of the stream at fixed detuning. In the one-atom coupling configuration, therefore, the exchange coupling acts primarily as a renormalization of the stream bias rather than as a source of explicit coherence-assisted enhancement.

**Appendix B. Additional limits and practical evaluation for the two-atom coupling configuration**

### B.1 Limiting cases and parameter trends

The steady-state cavity occupation for the two-atom coupling configuration is given in the main text by Eq. (42),

$$n^*(\Delta) = \frac{\kappa\bar{n}_1 + Rr_1^{(2)}\phi_2^2 L(\Delta)}{\kappa + R(r_2^{(2)} - r_1^{(2)})\phi_2^2 L(\Delta)}, \qquad (42)$$

with effective cavity temperature

$$T_{\text{cav}}(\Delta) = \frac{\hbar\omega_1}{k_B \ln(1 + 1/n^*(\Delta))} \qquad (43)$$

Several limiting cases are useful for interpreting the two-atom cooling mechanism.

First, in the resonant limit $\Delta = 0$, the overlap factor satisfies $L(0) = 1$, and the steady state reduces to

$$n^* = \frac{\kappa\bar{n}_1 + Rr_1^{(2)}\phi_2^2}{\kappa + R(r_2^{(2)} - r_1^{(2)})\phi_2^2}, \qquad (\text{B1})$$

with

$$T_{\text{cav}} = \frac{\hbar\omega_1}{k_B \ln(1 + 1/n^*)}. \qquad (\text{B2})$$

Thus, the detuned theory reduces smoothly to the resonant two-atom result when the pair and cavity are spectrally aligned.

Second, in the absence of the phonon bath, $\kappa \to 0$, the common overlap factor $L(\Delta)$cancels between numerator and denominator, and the cavity steady state becomes

$$n^*(\Delta) \underset{\kappa\to 0}{\to} \frac{r_1^{(2)}}{r_2^{(2)} - r_1^{(2)}}. \qquad (\text{B3})$$

In this limit,

$$\frac{n^*}{n^* + 1} = \frac{r_1^{(2)}}{r_2^{(2)}}, \qquad (\text{B4})$$

so the effective cavity temperature is determined entirely by the detailed balance imposed by the correlated pair stream. The important point is that this limit is independent of detuning: once the cavity bath is removed, detuning affects both the effective pump and effective loss by the same multiplicative factor, and therefore does not change the steady-state ratio.

Third, if the stream is effectively absent, either because $R\phi_2^2 \to 0$ or because $|\Delta| \gg \Gamma_{\text{over}}$ so that $L(\Delta) \to 0$, then

$$n^*(\Delta) \longrightarrow \bar{n}_1, \qquad \text{(B5)}$$

and the cavity simply thermalizes to the ambient bath temperature $T_{\text{bath}}$. This provides a useful reference point for distinguishing genuine reservoir-induced cooling from ordinary thermal equilibration.

A further practical trend concerns the interaction time $\tau$. Since $\phi_2^2 = \chi g^2 \tau^2$, increasing $\tau$strengthens the bare collision-induced pump and damping. At the same time, the detuning filter depends on the effective overlap width $\Gamma_{\text{over}} = \kappa + 1/\tau$, so that changing $\tau$ also reshapes the spectral overlap. For fixed $g$, $\Delta$, and $\kappa$, this produces a competition between stronger collisions and reduced bandwidth, implying that the net cooling or heating leverage of the stream is generally maximized at an intermediate $\tau$.

Finally, the roles of $\rho_{nd}$ and $\chi$ are distinct and complementary. The coherence $\rho_{nd}$ shifts both $r_1^{(2)}$ and $r_2^{(2)}$ by the same amount, thereby changing the absolute scale of the stream-induced pump and loss without modifying the net damping bias $r_2^{(2)} - r_1^{(2)}$. By contrast, $\chi$ is purely a geometric/phase parameter that rescales the collective interaction strength through $\phi_2^2 = \chi\phi^2$. The detuning filter $L(\Delta)$ then acts as a spectroscopic gate multiplying both effects. In this way, the two-atom configuration separates cleanly into three control layers: pair thermodynamics $(T_{\text{atom}}, \omega, \lambda)$, interaction geometry $(\chi)$, and spectral overlap $(\Delta, \tau, \kappa)$.

### B.2 Practical evaluation of the analytic steady state

For practical evaluation of the two-atom steady state, one first specifies the thermodynamic parameters of the incoming pair reservoir: the pair temperature $T_{\text{atom}}$, the local level splitting $\omega$, and the intra-pair exchange coupling $\lambda$. From these quantities one computes the partition function $Z$ and the pair-state matrix elements $\rho_e$, $\rho_g$, $\rho_d$, and $\rho_{nd}$ using Eqs. (2)–(3). The two-atom stream coefficients then follow directly from Eq. (30),

$$r_1^{(2)} = \rho_e + \rho_d + \rho_{nd}, \qquad r_2^{(2)} = \rho_g + \rho_d + \rho_{nd}. \qquad \text{(B6)}$$

Next, one fixes the cavity parameters $\omega_1$ and $T_{\text{bath}}$, which determine the thermal bath occupation

$$\bar{n}_1 = \frac{1}{\exp\left(\hbar\omega_1 / k_B T_{\text{bath}}\right) - 1}. \qquad \text{(B7)}$$

The collision parameters $g$ and $\tau$ then define the one-atom small angle $\phi = g\tau$, while the collective enhancement factor $\chi$ sets the effective two-atom strength $\phi_2^2 = \chi\phi^2$. The stream arrival rate $R$ and the cavity damping rate $\kappa$ complete the dynamical input.

To incorporate detuning, one specifies the atom–cavity mismatch

$$\Delta = \omega - \omega_1 \qquad \text{(B8)}$$

and evaluates the effective overlap width

$$\Gamma_{\text{over}} = \kappa + \frac{1}{\tau}, \qquad \text{(B9)}$$

together with the detuning filter

$$L(\Delta) = \frac{1}{1 + (2\Delta/\Gamma_{\text{over}})^2}. \qquad \text{(B10)}$$

At that stage all ingredients entering the two-atom rate equation are known.

The detuning-dressed damping and injection rates are then

$$\Gamma_{\downarrow}^{(2)}(\Delta) = \kappa + R(r_2^{(2)} - r_1^{(2)})\phi_2^2 L(\Delta), \qquad \text{(B11)}$$

$$J_{\uparrow}^{(2)}(\Delta) = \kappa\bar{n}_1 + R r_1^{(2)}\phi_2^2 L(\Delta). \qquad \text{(B12)}$$

Substituting these into Eq. (42) gives the steady-state cavity occupation in the compact form

$$n^*(\Delta) = \frac{J_{\uparrow}^{(2)}(\Delta)}{\Gamma_{\downarrow}^{(2)}(\Delta)} = \frac{\kappa\bar{n}_1 + R r_1^{(2)}\phi_2^2 L(\Delta)}{\kappa + R(r_2^{(2)} - r_1^{(2)})\phi_2^2 L(\Delta)}. \qquad \text{(B13)}$$

The corresponding effective cavity temperature is then obtained from Eq. (43),

$$T_{\text{cav}}(\Delta) = \frac{\hbar\omega_1}{k_B \ln(1 + 1/n^*(\Delta))}. \qquad \text{(B14)}$$

This workflow makes transparent how the different physical controls enter the theory. The parameters $(T_{\text{atom}}, \omega, \lambda)$ determine the pair-state thermodynamics and therefore the coefficients $r_1^{(2)}$ and $r_2^{(2)}$; the parameters $(g, \tau, \chi)$ determine the bare collision strength; the parameters $(\Delta, \Gamma_{\text{over}})$ determine the spectral-overlap penalty; and the parameters $(\kappa, T_{\text{bath}})$ determine the background thermalization toward the phonon bath. In the limit $\Delta \to 0$, one has $L(\Delta) \to 1$, so the resonant formulas are recovered directly. Conversely, for $|\Delta| \gg \Gamma_{\text{over}}$, the stream contribution is strongly suppressed and $n^*(\Delta)$ approaches $\bar{n}_1$. If one wishes to return to the one-atom coupling configuration while retaining the same bath and detuning treatment, it is sufficient to replace $r_1^{(2)}, r_2^{(2)} \to r_1, r_2$ and $\phi_2^2 \to \phi^2$; the overlap filter $L(\Delta)$ and the phonon-bath terms remain unchanged.

**Acknowledgement**

This study is partially based on work supported by AFOSR and LPS under contract numbers FA9550-23-1-0302 and FA9550-23-1-0763, and by the NSF under grant number CBET-2110603.

**References**

[1] Bluefors, "Dilution Refrigerator Measurement System", accessed November 28, 2025, https://bluefors.com/products/dilution-refrigerator-measurement-systems/ld-dilution-refrigerator-measurement-system/.

[2] Bluefors, LDHe Measurement System, accessed November 28, 2025, https://bluefors.com/products/1k-systems/ldhe-measurement-system/.

[3] Uhlig, Kurt. "Dry dilution refrigerator with 4He-1 K-loop." *Cryogenics* 66 (2015): 6-12.

[4] Bluefors, "Introducing the XLDHe High Power System - Ultimate Cooling for 1 K Experiments," accessed November 28, 2025, https://bluefors.com/news/introducing-the-xldhe-high-power-system-ultimate-cooling-for-1-k-experiments/

[5] Bluefors, "XLDHe High Power System," accessed November 28, 2025, https://bluefors.com/products/1k-systems/xldhe-high-power-system

[6] Bluefors, "Introducing the Ultra-Compact Dilution Refrigerator System", accessed November 28, 2025, https://bluefors.com/news/introducing-the-ultra-compact-dilution-refrigerator-system

[7] Bluefors, "LD Dilution Refrigerator Measurement System", accessed November 28, 2025, https://bluefors.com/products/dilution-refrigerator-measurement-systems/ld-dilution-refrigerator-measurement-system

[8] Krantz, Philip, Morten Kjaergaard, Fei Yan, Terry P. Orlando, Simon Gustavsson, and William D. Oliver. "A quantum engineer's guide to superconducting qubits." *Applied physics reviews* 6, no. 2 (2019).

[9] Blais, Alexandre, Arne L. Grimsmo, Steven M. Girvin, and Andreas Wallraff. "Circuit quantum electrodynamics." *Reviews of Modern Physics* 93, no. 2 (2021): 025005.

[10] Gardiner, Crispin W., and Matthew J. Collett. "Input and output in damped quantum systems: Quantum stochastic differential equations and the master equation." *Physical Review A* 31, no. 6 (1985): 3761.

[11] Ciccarello, Francesco, Salvatore Lorenzo, Vittorio Giovannetti, and G. Massimo Palma. "Quantum collision models: Open system dynamics from repeated interactions." *Physics Reports* 954 (2022): 1-70.

[12] Clerk, Aashish A. "Quantum noise and quantum measurement." Quantum Machines: Measurement and Control of Engineered Quantum Systems (2008).
[13] Filipowicz, P., Javanainen, J., & Meystre, P., Theory of a microscopic maser, Phys. Rev. A 34, 3077 (1986)
[14] Linden, Noah, Sandu Popescu, and Paul Skrzypczyk. "How small can thermal machines be? The smallest possible refrigerator." *Physical review letters* 105, no. 13 (2010): 130401.
[15] Correa, Luis A., José P. Palao, Daniel Alonso, and Gerardo Adesso. "Quantum-enhanced absorption refrigerators." *Scientific reports* 4, no. 1 (2014): 3949.
[16] Bhandari, Bibek, and Andrew N. Jordan. "Minimal two-body quantum absorption refrigerator." *Physical Review B* 104, no. 7 (2021): 075442.
[17] Englert, Berthold-Georg. "Elements of micromaser physics." *arXiv preprint quant-ph/0203052* (2002).
[18] Walls, D. F., and Gerard J. Milburn. "Quantum Optics and Quantum Foundations." In *Quantum Optics*, pp. 225-244. Cham: Springer Nature Switzerland, 2025.
[19] Dillenschneider, Raoul, and Eric Lutz. "Energetics of quantum correlations." *Europhysics Letters* 88, no. 5 (2009): 50003.

[20] Zhang, Yuan, Qilong Wu, Hao Wu, Xun Yang, Shi-Lei Su, Chongxin Shan, and Klaus Mølmer. "Microwave mode cooling and cavity quantum electrodynamics effects at room temperature with optically cooled nitrogen-vacancy center spins." *npj Quantum Information* 8, no. 1 (2022): 125.

[21] Klaus Hepp, Elliott H. Lieb, On the Superradiant Phase Transition for Molecules in a Quantized Radiation Field: the Dicke Maser Model, ANNALS OF PHYSICS 76, 360-404 (1973)

[22] Fahey, Donald P., Kurt Jacobs, Matthew J. Turner, Hyeongrak Choi, Jonathan E. Hoffman, Dirk Englund, and Matthew E. Trusheim. "Steady-state microwave mode cooling with a diamond N-V ensemble." *Physical Review Applied* 20, no. 1 (2023): 014033.

[23] Magnard, Paul, Philipp Kurpiers, Baptiste Royer, Theo Walter, J-C. Besse, Simone Gasparinetti, Marek Pechal et al. "Fast and unconditional all-microwave reset of a superconducting qubit." *Physical review letters* 121, no. 6 (2018): 060502.
[24] Ristè, D., C. C. Bultink, Konrad W. Lehnert, and L. DiCarlo. "Feedback control of a solid-state qubit using high-fidelity projective measurement." *Physical review letters* 109, no. 24 (2012): 240502.
[25] Shankar, S., M. Hatridge, Z. Leghtas, K. M. Sliwa, A. Narla, U. Vool, S. M. Girvin, L. Frunzio, M. Mirrahimi, and M. H. Devoret. "Stabilizing entanglement autonomously between two superconducting qubits." *arXiv preprint arXiv:1307.4349* (2013).
[26] Gu, Xu-Yang, Da'er Feng, Zhen-Yu Peng, Gui-Han Liang, Yang He, Yongxi Xiao, Ming-Chuan Wang et al. "Engineering a Multi-Mode Purcell Filter for Superconducting-Qubit Reset and Readout with Intrinsic Purcell Protection." *arXiv preprint arXiv:2507.04676* (2025).
[27] Sunada, Yoshiki, Shingo Kono, Jesper Ilves, Shuhei Tamate, Takanori Sugiyama, Yutaka Tabuchi, and Yasunobu Nakamura. "Fast readout and reset of a superconducting qubit coupled to a resonator with an intrinsic Purcell filter." *Physical Review Applied* 17, no. 4 (2022): 044016.
[28] Ding, Jiayu, Yulong Li, He Wang, Guangming Xue, Tang Su, Chenlu Wang, Weijie Sun et al. "Multipurpose architecture for fast reset and protective readout of superconducting qubits." *Physical Review Applied* 23, no. 1 (2025): 014012.
[29] Reed, Matthew D., Blake R. Johnson, Andrew A. Houck, Leonardo DiCarlo, Jerry M. Chow, David I. Schuster, Luigi Frunzio, and Robert J. Schoelkopf. "Fast reset and suppressing spontaneous emission of a superconducting qubit." *Applied Physics Letters* 96, no. 20 (2010).
[30] Alhambra, Álvaro M., Matteo Lostaglio, and Christopher Perry. "Heat-bath algorithmic cooling with optimal thermalization strategies." *Quantum* 3 (2019): 188.
[31] Underwood, Devin, Joseph A. Glick, Ken Inoue, David J. Frank, John Timmerwilke, Emily Pritchett, Sudipto Chakraborty et al. "Using cryogenic CMOS control electronics to enable a two-qubit cross-resonance gate." *PRX Quantum* 5, no. 1 (2024): 010326.
[32] Pellerano, Stefano, Sushil Subramanian, Jong-Seok Park, Bishnu Patra, Todor Mladenov, Xiao Xue, Lieven MK Vandersypen, Masoud Babaie, Edoardo Charbon, and Fabio Sebastiano. "Cryogenic CMOS for qubit control and readout." In *2022 IEEE Custom Integrated Circuits Conference (CICC)*, pp. 01-08. IEEE, 2022.

[33] Gross, Michel, and Serge Haroche. "Superradiance: An essay on the theory of collective spontaneous emission." *Physics reports* 93, no. 5 (1982): 301-396.
[34] Dicke, Robert H. "Coherence in spontaneous radiation processes." *Physical review* 93, no. 1 (1954): 99.
[35] Poyatos, J. F., J. Ignacio Cirac, and P. Zoller. "Quantum reservoir engineering with laser cooled trapped ions." *Physical review letters* 77, no. 23 (1996): 4728.
[36] Forn-Díaz, P., L. Lamata, E. Rico, J. Kono, and E. Solano. "Ultrastrong coupling regimes of light-matter interaction." *Reviews of Modern Physics* 91, no. 2 (2019): 025005.
[37] Mundada, Pranav, Gengyan Zhang, Thomas Hazard, and Andrew Houck. "Suppression of qubit crosstalk in a tunable coupling superconducting circuit." *Physical Review Applied* 12, no. 5 (2019): 054023.
[38] Campbell, Daniel L., Archana Kamal, Leonardo Ranzani, Michael Senatore, and Matthew D. LaHaye. "Modular tunable coupler for superconducting circuits." *Physical Review Applied* 19, no. 6 (2023): 064043.
[39] Ding, Leon, Max Hays, Youngkyu Sung, Bharath Kannan, Junyoung An, Agustin Di Paolo, Amir H. Karamlou et al. "High-fidelity, frequency-flexible two-qubit fluxonium gates with a transmon coupler." *Physical Review X* 13, no. 3 (2023): 031035.
[40] Gargiulo, O., S. Oleschko, J. Prat-Camps, M. Zanner, and G. Kirchmair. "Fast flux control of 3D transmon qubits using a magnetic hose." *Applied Physics Letters* 118, no. 1 (2021).
[41] Han, Lian-Chen, Yu Xu, Jin Lin, Fu-Sheng Chen, Shao-Wei Li, Cheng Guo, Na Li et al. "Active reset of superconducting qubits using the electronics based on RF switches." *AIP Advances* 13, no. 9 (2023).
[42] Shah, Parth S., Frank Yang, Chaitali Joshi, and Mohammad Mirhosseini. "Stabilizing remote entanglement via waveguide dissipation." *PRX Quantum* 5, no. 3 (2024): 030346.
[43] Li, Ziqian, Tanay Roy, Yao Lu, Eliot Kapit, and David I. Schuster. "Autonomous stabilization with programmable stabilized state." *Nature Communications* 15, no. 1 (2024): 6978.
[44] Zhou, Yu, Zhenxing Zhang, Zelong Yin, Sainan Huai, Xiu Gu, Xiong Xu, Jonathan Allcock et al. "Rapid and unconditional parametric reset protocol for tunable superconducting qubits." *Nature Communications* 12, no. 1 (2021): 5924.
[45] Sirois, Adam J., M. A. Castellanos-Beltran, M. P. DeFeo, L. Ranzani, F. Lecocq, R. W. Simmonds, J. D. Teufel, and J. Aumentado. "Coherent-state storage and retrieval between superconducting cavities using parametric frequency conversion." *Applied Physics Letters* 106, no. 17 (2015).
[46] Chapman, Benjamin J., Stijn J. De Graaf, Sophia H. Xue, Yaxing Zhang, James Teoh, Jacob C. Curtis, Takahiro Tsunoda et al. "High-on-off-ratio beam-splitter interaction for gates on bosonically encoded qubits." *PRX Quantum* 4, no. 2 (2023): 020355.
[47] Serafini, A., M. Lostaglio, S. Longden, U. Shackerley-Bennett, C-Y. Hsieh, and G. Adesso. "Gaussian thermal operations and the limits of algorithmic cooling." *Physical Review Letters* 124, no. 1 (2020): 010602.
[48] Lvov, Dmitrii S., Sergei A. Lemziakov, Elias Ankerhold, Joonas T. Peltonen, and Jukka P. Pekola. "Thermometry based on a superconducting qubit." *Physical Review Applied* 23, no. 5 (2025): 054079.
[49] Ronzani, Alberto, Bayan Karimi, Jorden Senior, Yu-Cheng Chang, Joonas T. Peltonen, ChiiDong Chen, and Jukka P. Pekola. "Tunable photonic heat transport in a quantum heat valve." *Nature Physics* 14, no. 10 (2018): 991-995.
[50] Huang, Jordan, Thomas J. DiNapoli, Gavin Rockwood, Ming Yuan, Prathyankara Narasimhan, Eesh Gupta, Mustafa Bal et al. "Fast Sideband Control of a Multimode Cavity Memory with Weak Dispersive Coupling to a Transmon." *Physical Review X* 16, no. 1 (2026): 011058.
[51] Vashaee, Daryoosh, and Jahanfar Abouie. "Microwave-induced cooling in double quantum dots: Achieving millikelvin temperatures to reduce thermal noise around spin qubits." *Physical Review B* 111, no. 1 (2025): 014305.
[52] Stockklauser, Anna, Pasquale Scarlino, Jonne V. Koski, Simone Gasparinetti, Christian Kraglund Andersen, Christian Reichl, Werner Wegscheider, Thomas Ihn, Klaus Ensslin, and Andreas Wallraff. "Strong coupling cavity QED with gate-defined double quantum dots enabled by a high impedance resonator." *Physical Review X* 7, no. 1 (2017): 011030.
[53] Mi, Xiao, J. V. Cady, D. M. Zajac, P. W. Deelman, and Jason R. Petta. "Strong coupling of a single electron in silicon to a microwave photon." *Science* 355, no. 6321 (2017): 156-158.
[54] Cai, Weizhou, Yuwei Ma, Weiting Wang, Chang-Ling Zou, and Luyan Sun. "Bosonic quantum error correction codes in superconducting quantum circuits." *Fundamental Research* 1, no. 1 (2021): 50-67.

[55] Lachance-Quirion, Dany, Marc-Antoine Lemonde, Jean Olivier Simoneau, Lucas St-Jean, Pascal Lemieux, Sara Turcotte, Wyatt Wright et al. "Autonomous quantum error correction of Gottesman-Kitaev-Preskill states." *Physical Review Letters* 132, no. 15 (2024): 150607.